\documentclass{article}
\usepackage{amsmath,graphicx}
\usepackage{dsfont}
\usepackage[title]{appendix}
\usepackage{authblk}

\usepackage{subfig}
\usepackage{dsfont}
\usepackage{float}
\usepackage{amsthm}

\title{Gradient-based Filter Design for the Dual-tree Wavelet Transform}

\author{Daniel Recoskie\thanks{Thanks to NSERC for funding.}~}
\author{Richard Mann}
\affil{University of Waterloo\\
	Cheriton School of Computer Science\\
	Waterloo, Canada\\
	\{dprecosk, mannr\}@uwaterloo.ca}
\date{}

\begin{document}

\maketitle

\begin{abstract}
The wavelet transform has seen success when incorporated into neural network architectures, such as in wavelet scattering networks. More recently, it has been shown that the dual-tree complex wavelet transform can provide better representations than the standard transform. With this in mind, we extend our previous method for learning filters for the 1D and 2D wavelet transforms into the dual-tree domain. We show that with few modifications to our original model, we can learn directional filters that leverage the properties of the dual-tree wavelet transform. 
\end{abstract}

\section{Introduction}
In this work we explore the task of learning filters for the dual-tree complex wavelet transform \cite{kingsbury1998dual,selesnick2005dual}. This transform was introduced to address several shortcomings of the separable, real-valued wavelet transform algorithm. However, the dual-tree transform requires greater care when designing filters. The added complexity makes the transform a good candidate to replace the traditional filter derivations with learning. We demonstrate that it is possible to learn filters for the dual-tree complex wavelet transform in a similar fashion to \cite{recoskie2018alearning,recoskie2018blearning}. We show that very few changes to the original autoencoder framework are necessary to learn filters that overcome the limitations of the separable 2D wavelet transform. 

Wavelet representations have been shown to perform well on a variety of machine learning tasks. Specifically, wavelet scattering networks have shown state-of-the-art results despite the fact they use a fixed representation (in contrast to the learned representations of convolutional neural networks) \cite{bruna2011classification,mallat2012group,bruna2013invariant,mallat2016understanding}. Similar work has been done using oriented bandpass filters in the SOE-Net \cite{hadji2017spatiotemporal}. More recently, it has been shown that extending this work using the dual-tree complex wavelet transform can lead to improved results \cite{singh2017dual,singh2017efficient}. In all of these examples, the filters used in the networks are fixed. Building upon our work in \cite{recoskie2018alearning,recoskie2018blearning}, the goal of this work is to demonstrate that it is possible to learn valid filters for the dual-tree complex wavelet transform for use in neural network architectures.

\section{Problems with the Wavelet Transform}
The discrete real wavelet transform has many desirable properties, such as a linear time algorithm and basis functions (wavelets) that are not fixed. However, the transform does have some drawbacks. The four major limitations that we will consider are: shift variance, oscillations, lack of directionality, and aliasing. In order to overcome these problems, we will need to make use of complex wavelets. We will restrict ourselves to a single formulation of the complex wavelet transform known as the dual-tree complex wavelet transform (DTCWT) \cite{kingsbury1998dual}. The DTCWT overcomes the limitations of the standard wavelet transform (with some computational overhead). Before going into the details of the DTCWT, we will discuss the issues with the standard wavelet transform. A more detailed discussion can be found in \cite{selesnick2005dual}.

\subsection{Shift Variance}
A major problem with the real-valued wavelet transform is shift variance. A desirable property for any representation is that small perturbations of the input should result only in small perturbations in the feature representation. In the case of the wavelet transform, translating the input (even by a single sample) can result in large changes to the wavelet coefficients. Figure \ref{fig:shifts} demonstrates the issue of shift variance. We plot the third scale wavelet coefficients of a signal composed of a single step edge. The signal is shifted by a single sample, and the coefficients are recomputed. Note that the coefficient values change significantly after the shift.

\subsection{Oscillations}
\begin{figure}[tb]
\centering
\subfloat[]{\includegraphics[width=.295\textwidth]{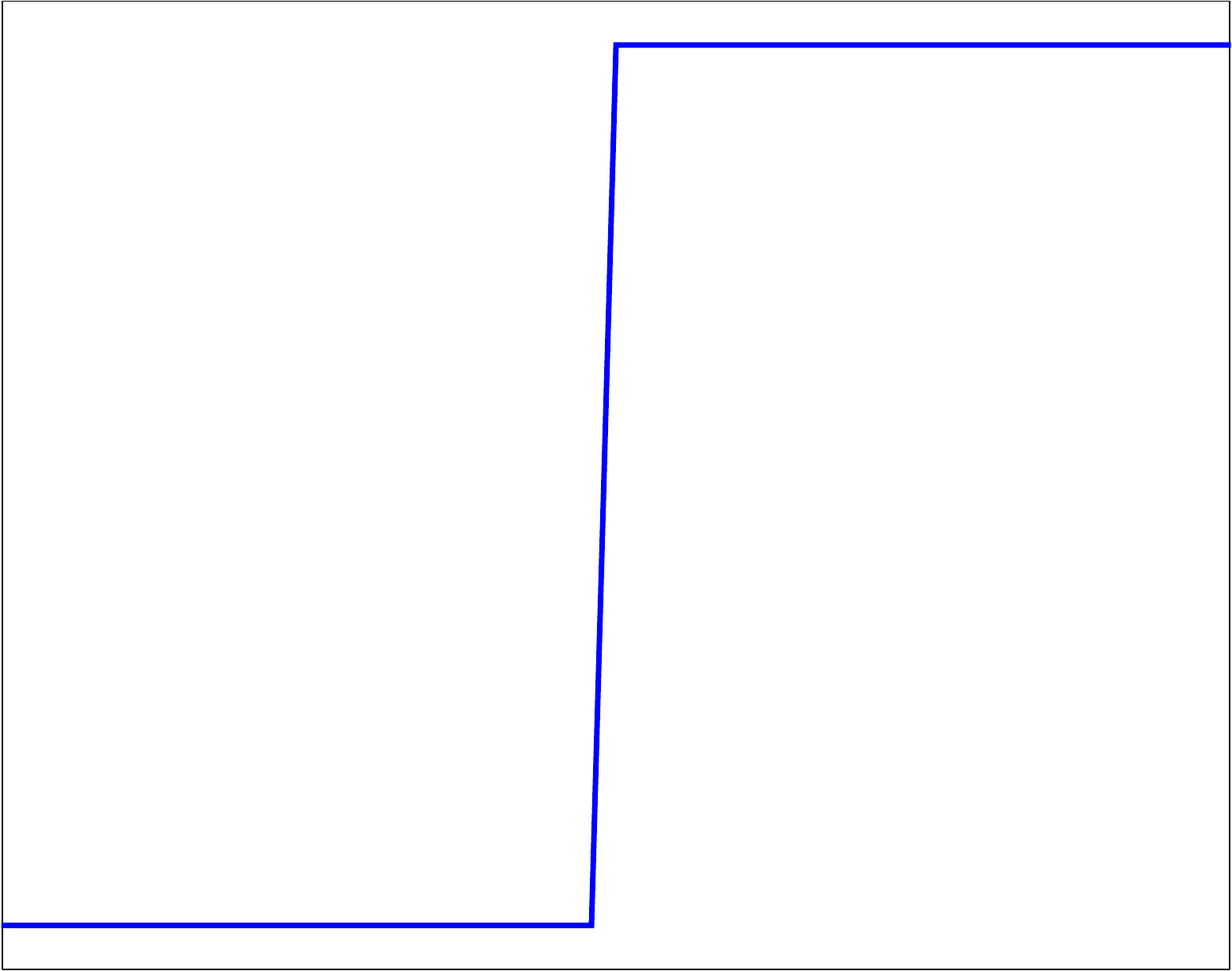}}
\hfill
\subfloat[]{\includegraphics[width=.3\textwidth]{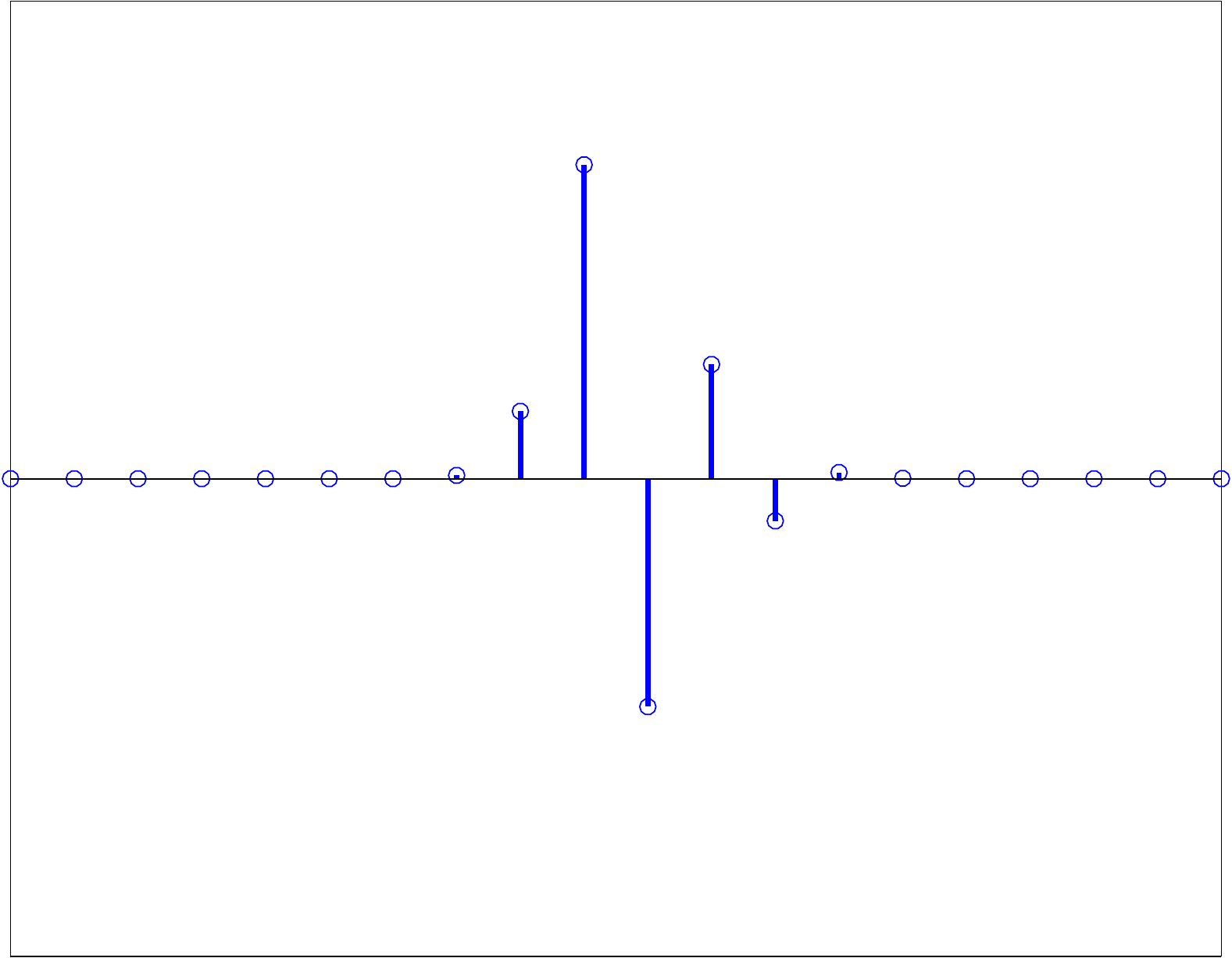}}
\hfill
\subfloat[]{\includegraphics[width=.3\textwidth]{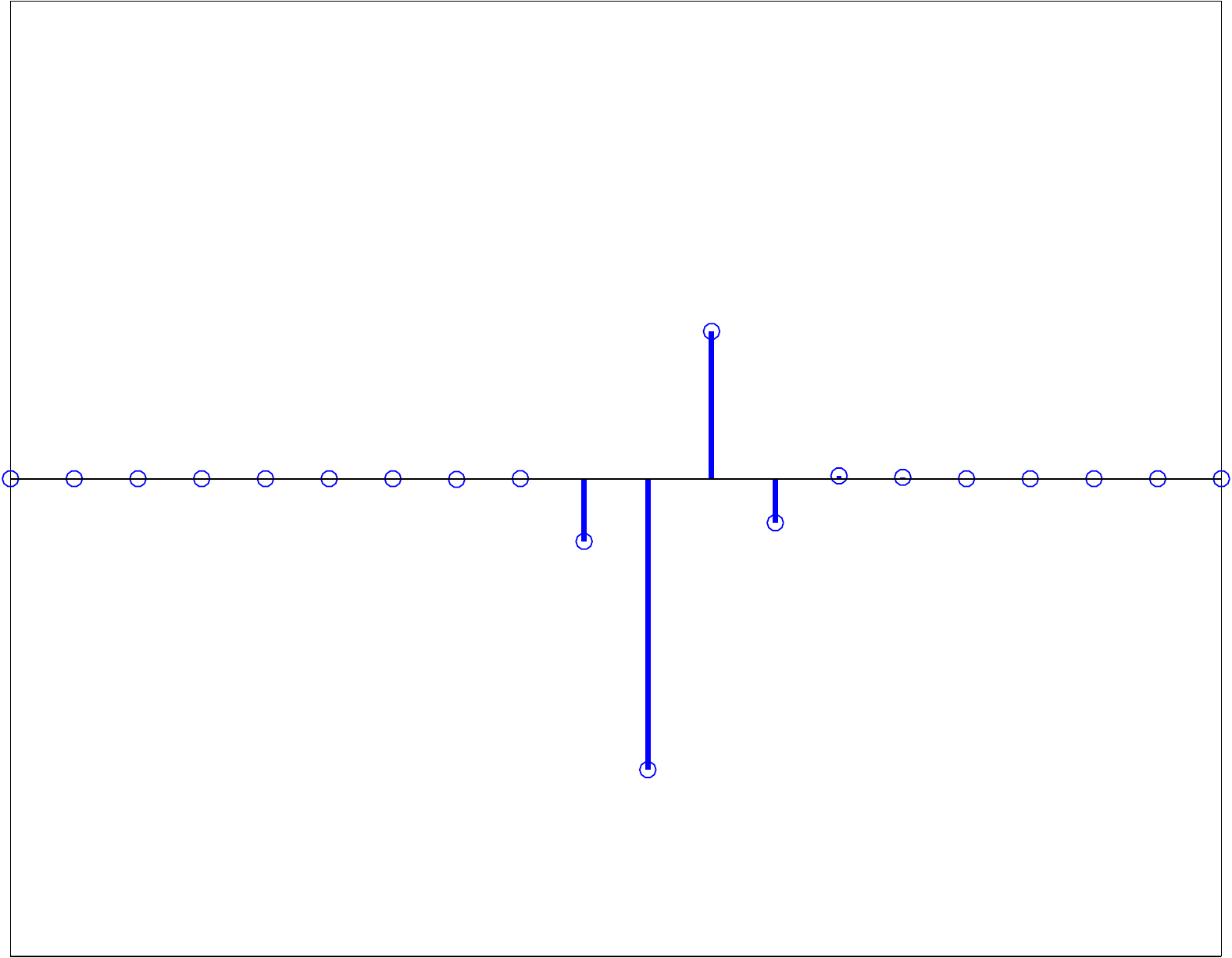}}
\caption{(a) A signal containing a single step edge. (b) The third level Daubechies 6 wavelet coefficients of (a). (c) Same as (b) but with the signal in (a) shifted by a single sample.}
\label{fig:shifts}
%
\centering
\subfloat[]{\includegraphics[height=.3\textwidth]{./oscilation-signal}}
\hfil
\subfloat[]{\includegraphics[height=.3\textwidth]{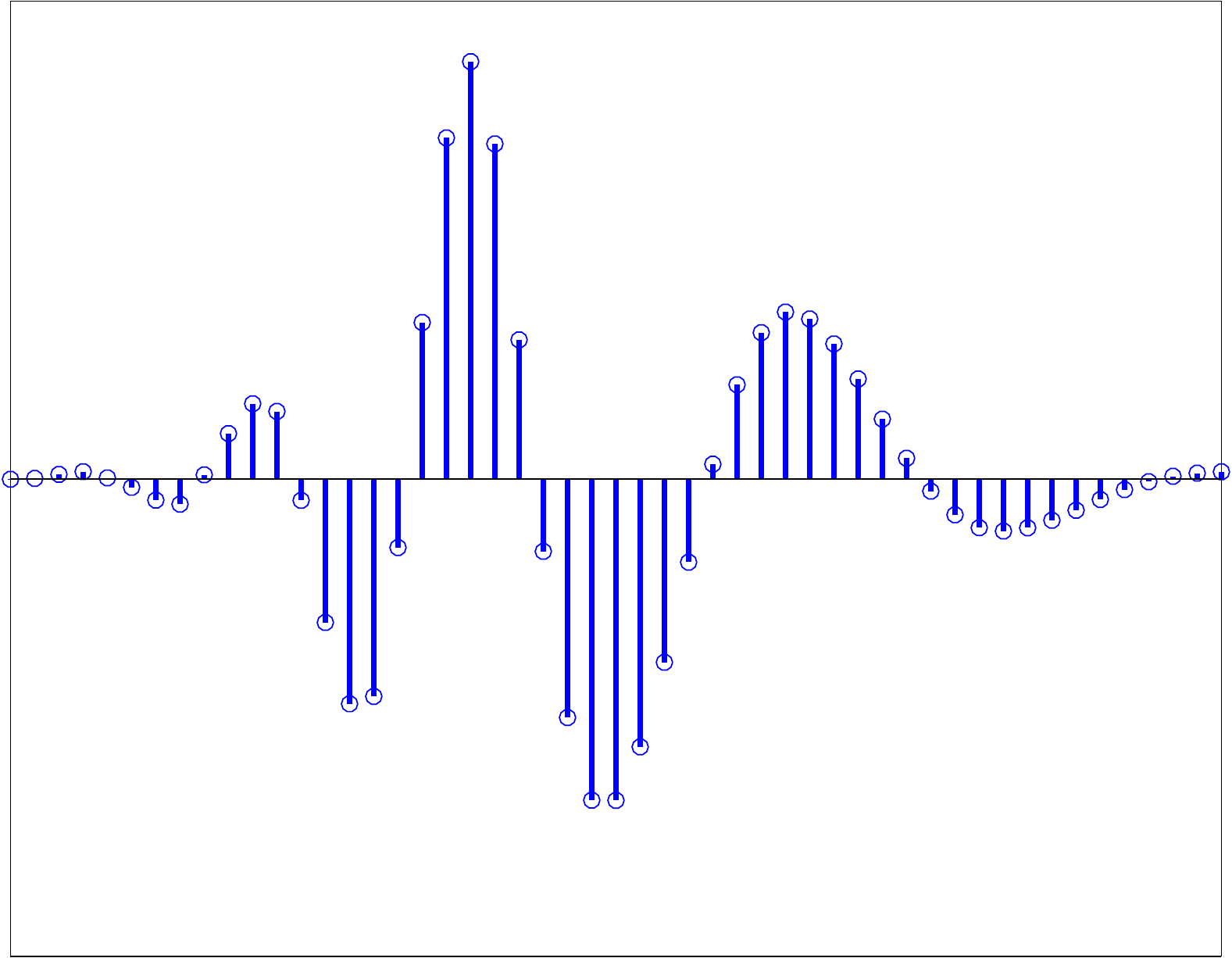}}
\caption{(a) A signal containing a single step edge. (b) The (undecimated) third level Daubechies 6 wavelet coeficients corresponding to the edge.}
\label{fig:oscillations}
\end{figure}

Figure \ref{fig:shifts} also demonstrates that wavelet coefficients are not stable near signal singularities such as step edges or impulses. Note how the wavelet coefficients change in value and sign near the edge. We can see this problem is more detail by looking at the non-decimated wavelet coefficients of the step edge in Figure \ref{fig:oscillations}. The coefficients necessarily oscillate because the wavelet filter is highpass.

\subsection{Lack of Directionality}
\begin{figure}[tb]
\centering
\includegraphics[width=1\textwidth]{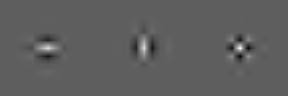}
\caption{Typical impulse response of filters used in the 2D wavelet transform.}
\label{fig:oriented}
%
\centering
\subfloat[]{
\includegraphics[width=.22\textwidth,trim={17cm 17cm 0 0},clip]{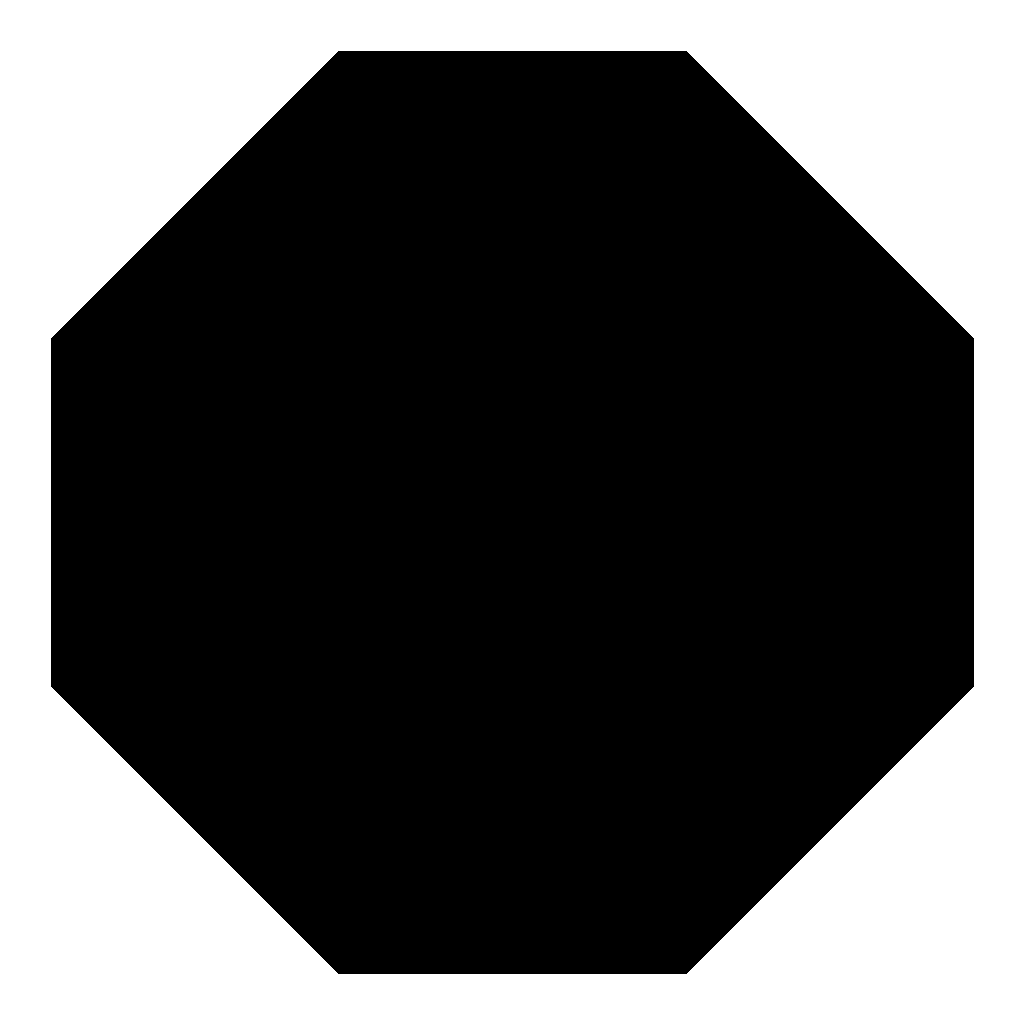}
\hfill
\includegraphics[width=.22\textwidth,trim={17cm 17cm 0 0},clip]{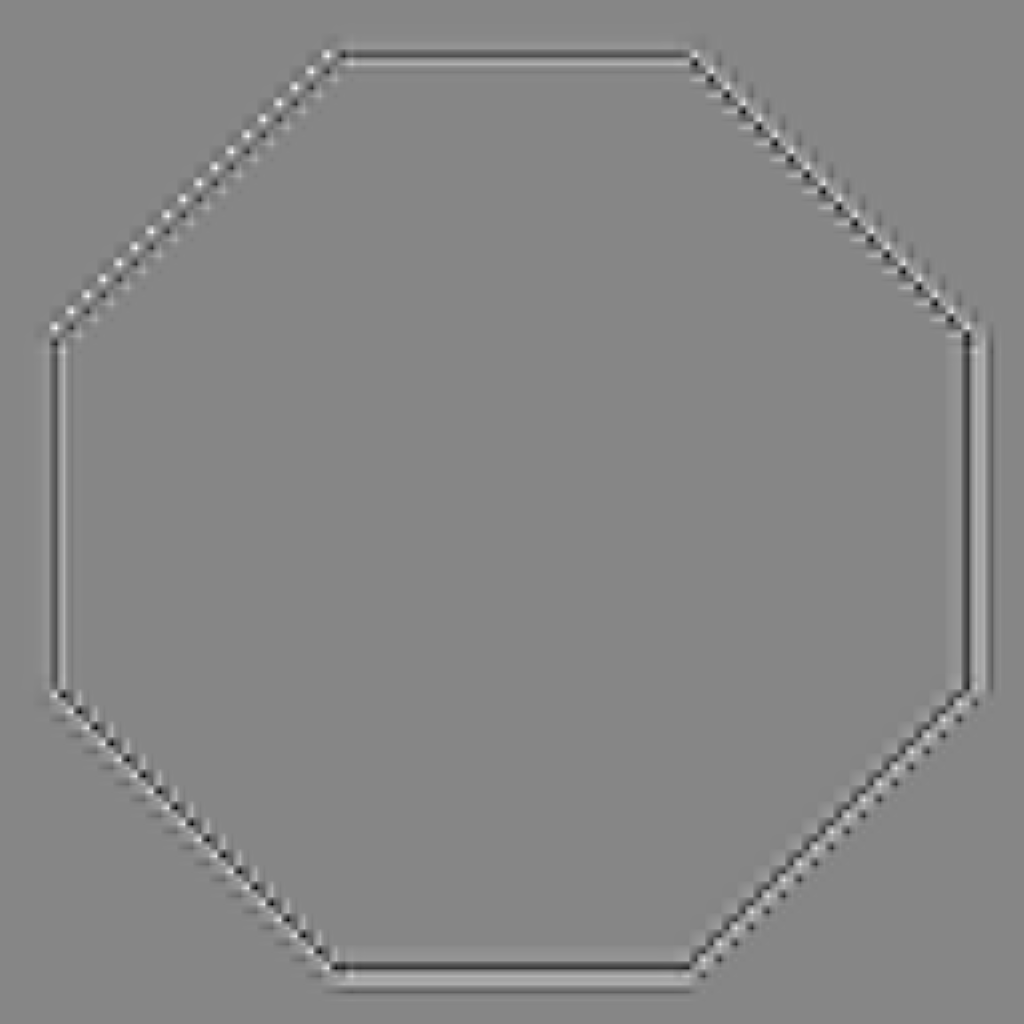}}
\hfill
\subfloat[]{
\includegraphics[width=.22\textwidth,trim={17cm 17cm 0 0},clip]{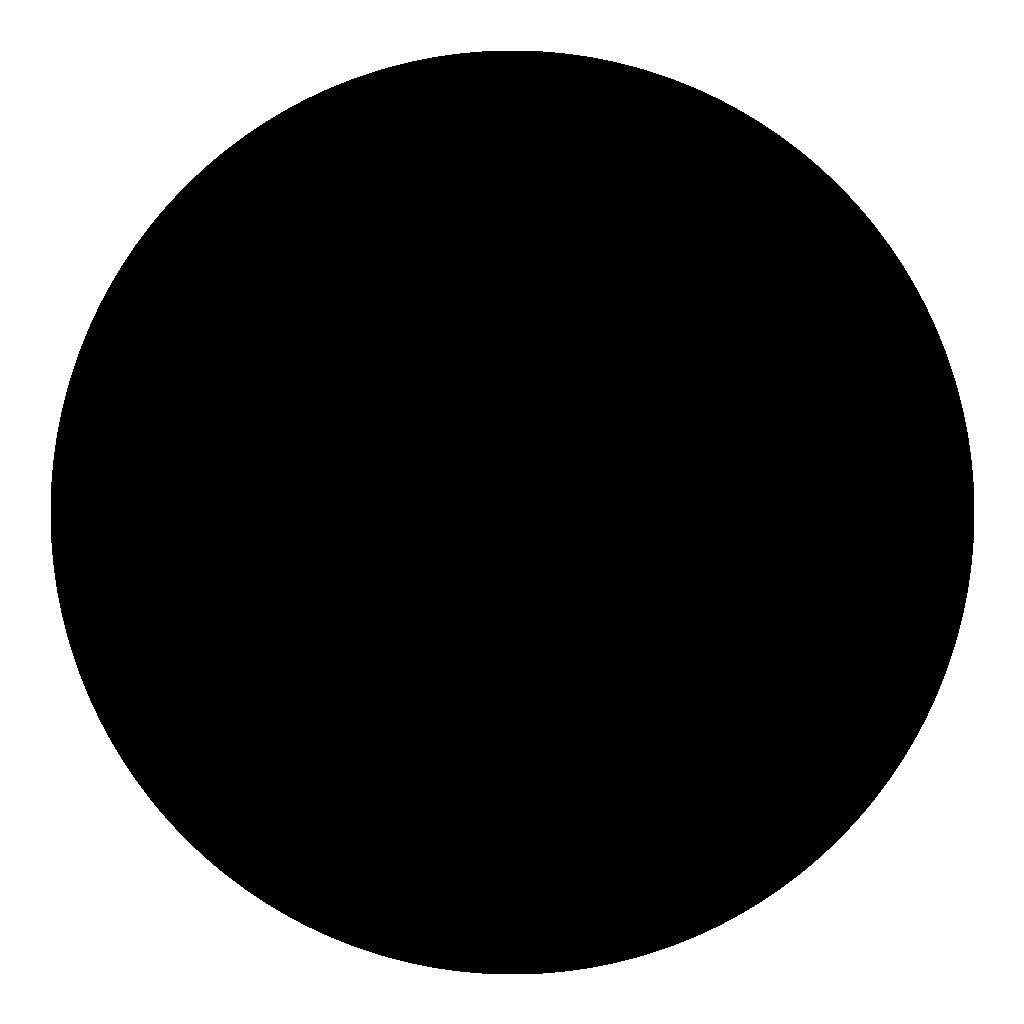}
\hfill
\includegraphics[width=.22\textwidth,trim={17cm 17cm 0 0},clip]{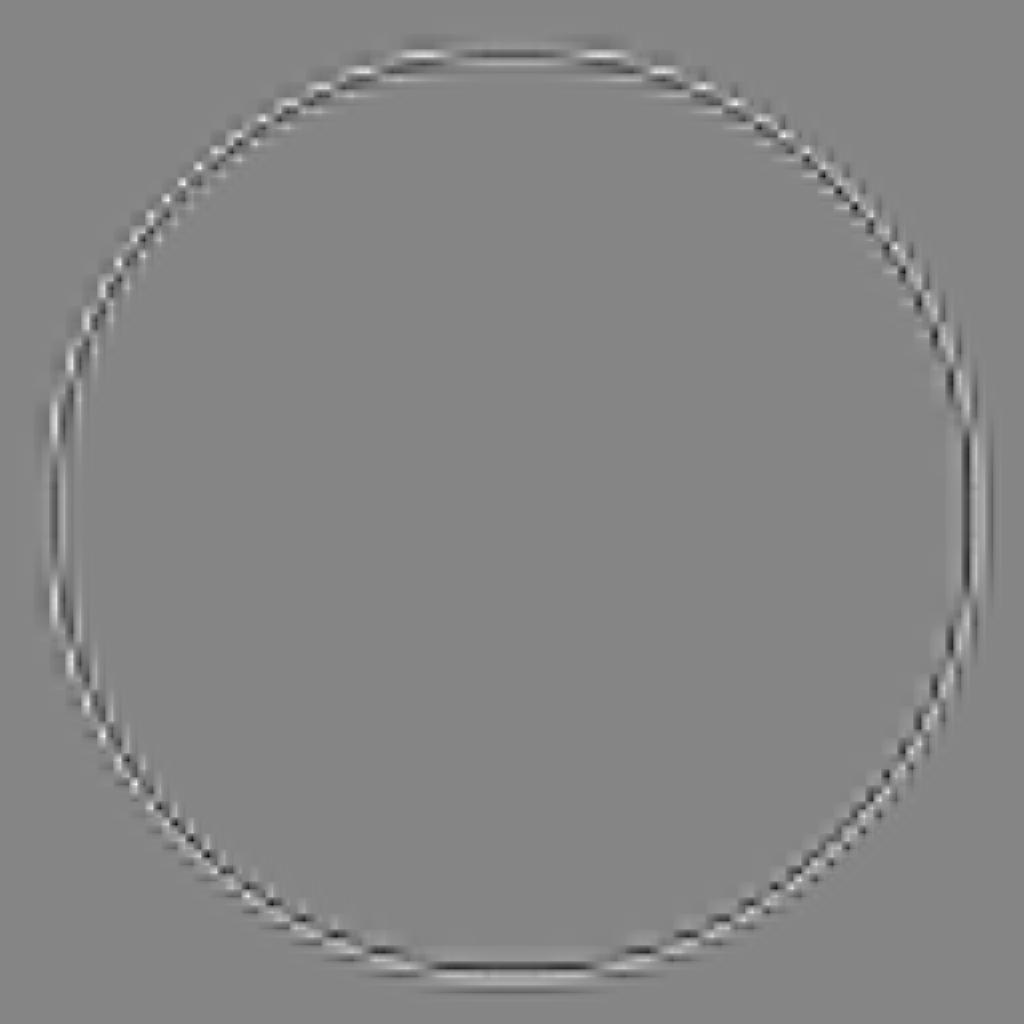}}
\caption{(a) Left: An image with three edge orientations. Right: Reconstruction of the image using only the fourth band Daubechies 2 wavelet coefficients. Note the irregularities on the edges that are not axis-aligned. (b) Same as (a), but using an image with a curved edge.}
\label{fig:bad-edges}
\end{figure}

The standard multidimensional wavelet transform algorithm makes use of separable filters. In other words, a single one-dimensional filter is applied along each dimension of the input. Though this method lends itself to an efficient implementation of the algorithm, there are some problems in the impulse responses of the filters. Namely, we can only properly achieve horizontal and vertical directions of the filters, while the diagonal direction suffers from checkerboard artifacts. Figure \ref{fig:oriented} shows impulse responses of a typical wavelet filter in the 2D transform. The impulse responses demonstrate these problems.

The problem of not having directionality can be seen by reconstructing simple images from wavelet coefficients at a single scale. Figure \ref{fig:bad-edges} shows two images reconstructed from their fourth scale wavelet coefficients. Note that the horizontal and vertical edges in the first image appear without any artifacts. The diagonal edges, on the other hand, have significant irregularities. The curved image illustrates this further, as almost all edges are off-axis.

\subsection{Aliasing}
\begin{figure}[tb]
\centering
\subfloat[]{\includegraphics[width=.3\textwidth]{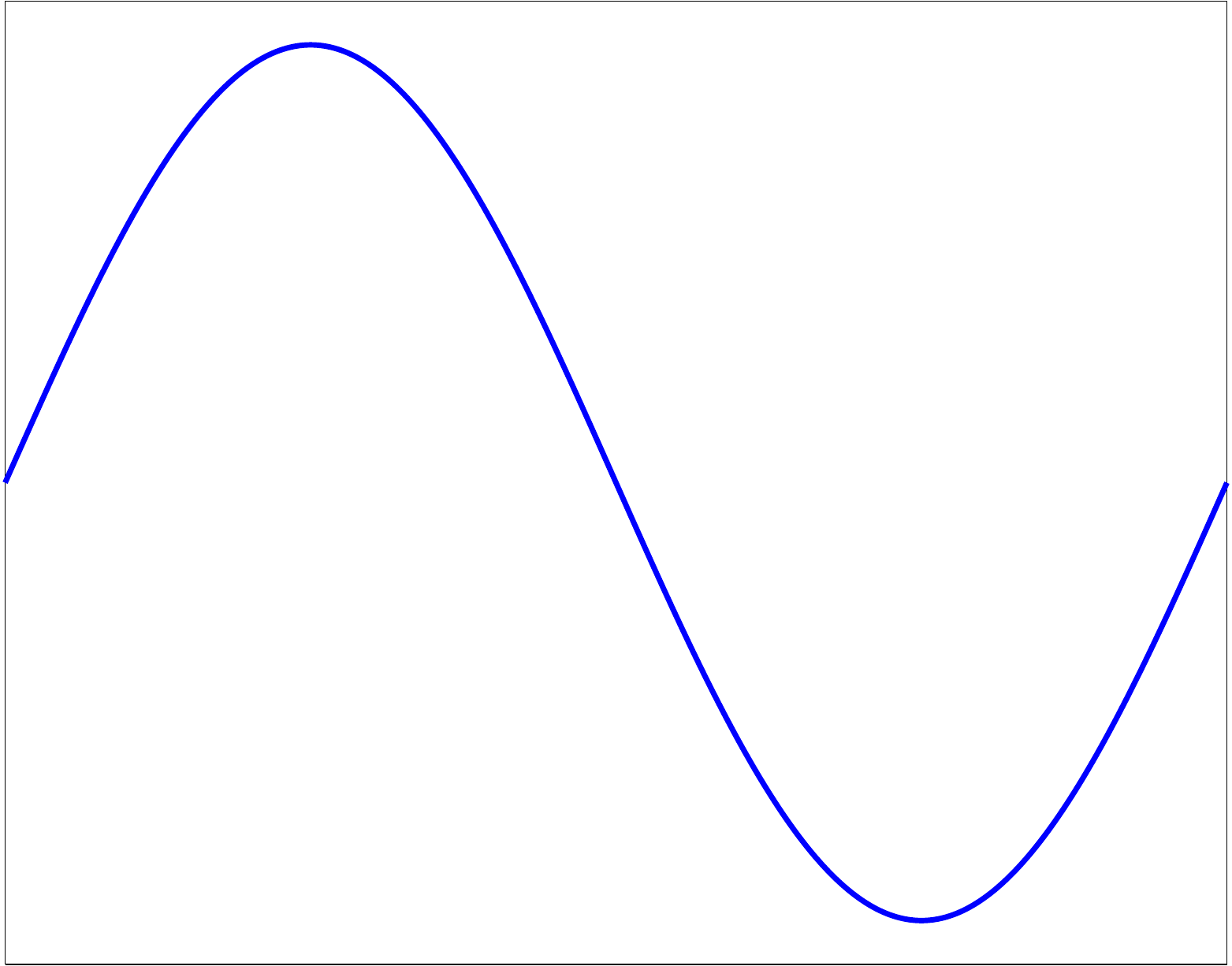}}
\hfill
\subfloat[]{\includegraphics[width=.3\textwidth]{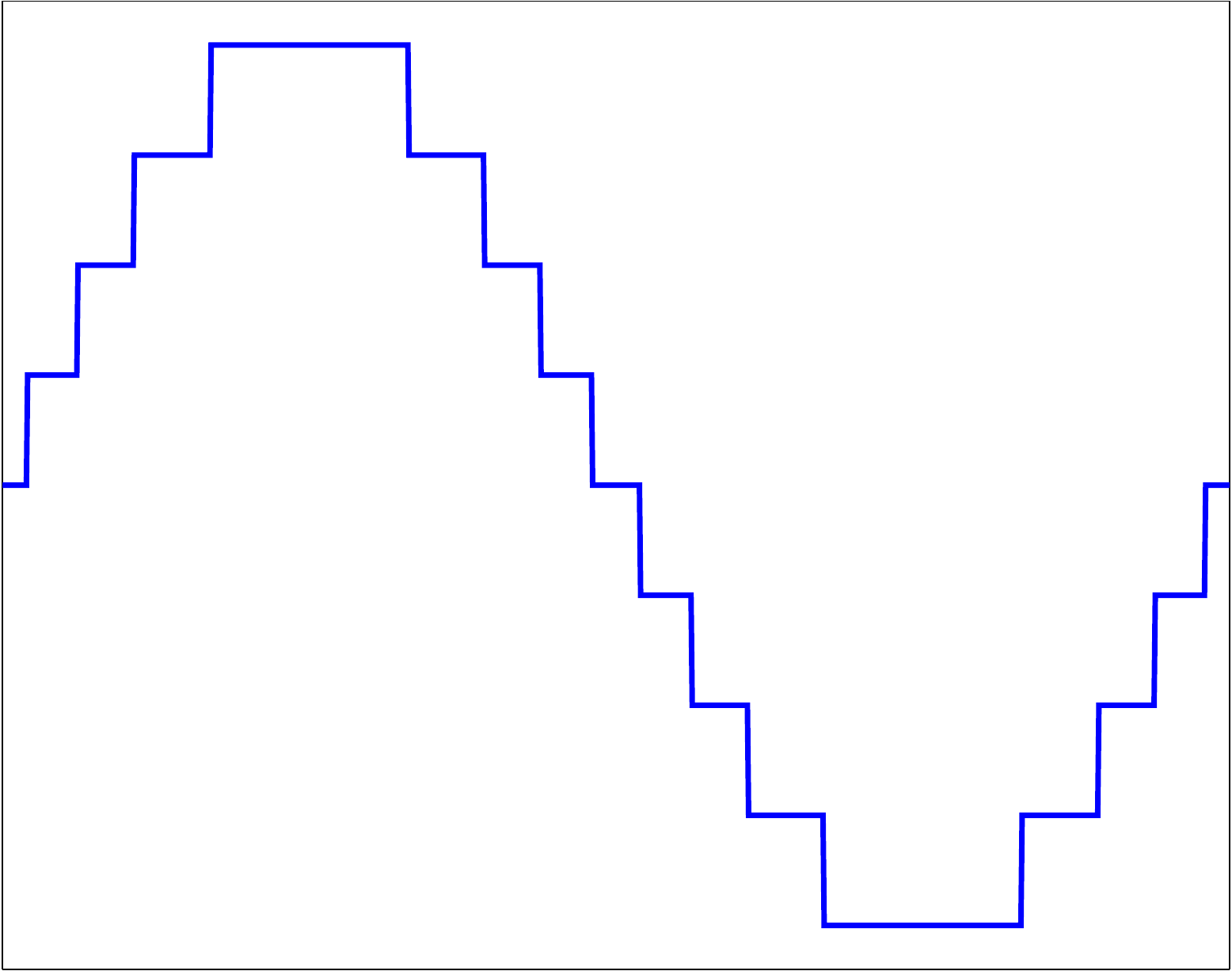}}
\hfill
\subfloat[]{\includegraphics[width=.3\textwidth]{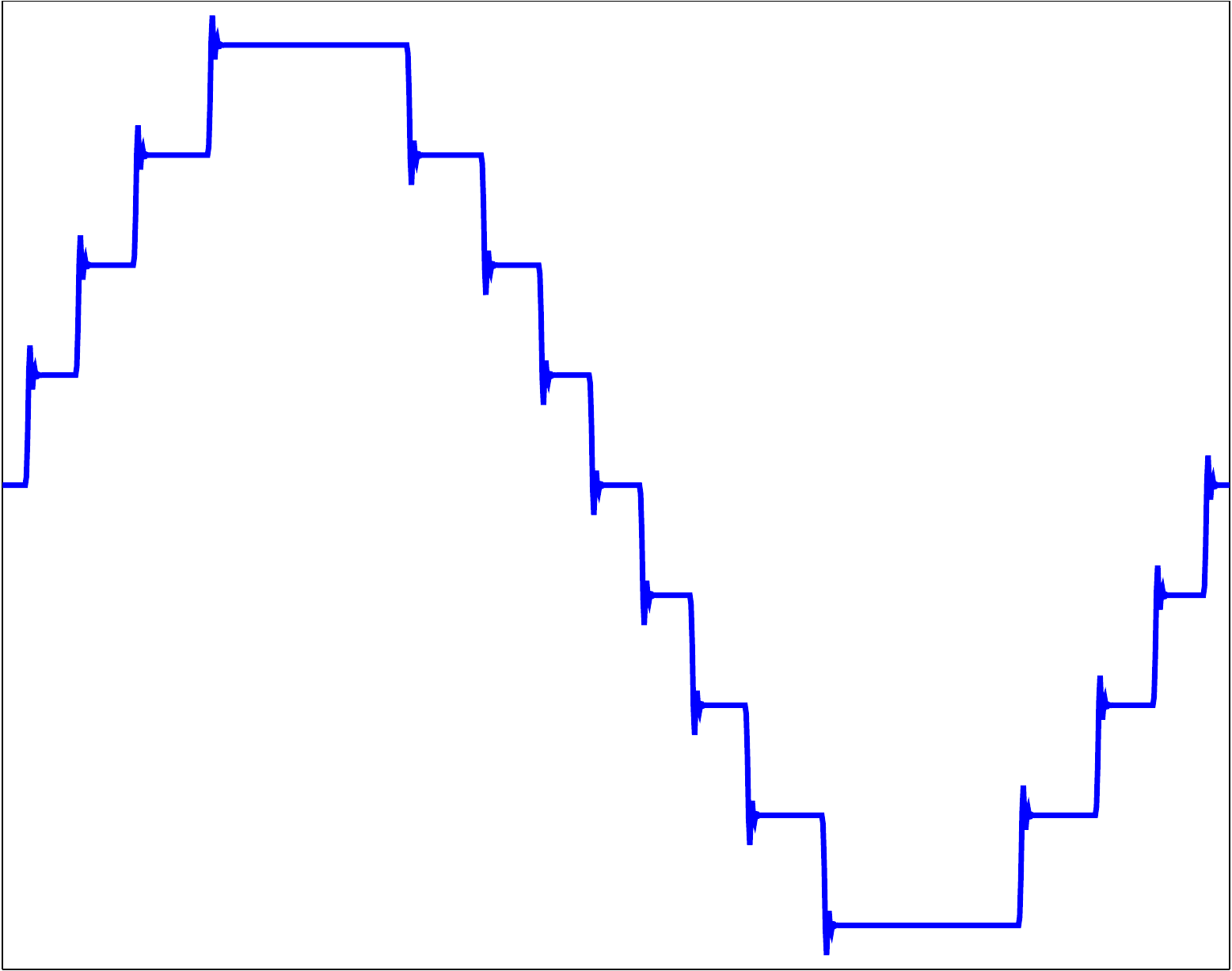}}
\caption{(a) Original signal. (b) The signal from (a) quantized to nine levels. (c) A one level wavelet transform is first applied to (a), the wavelet coefficients are quantized to nine levels, and an inverse transform is applied. Note the artifacts near the step edges.}
\label{fig:aliasing}
\end{figure}

The final problem we will discuss is aliasing. The standard decimating discrete wavelet transform algorithm downsamples the wavelet coefficients by a factor of two after the wavelet filters are applied. Normally we must apply a lowpass (antialiasing) filter prior to downsampling a signal  to prevent aliasing. Aliasing occurs when downsampling is performed on a signal that contains frequencies above the Nyquist rate. The energy of these high frequencies is reflected onto the low frequency components, causing artifacts. The wavelet transform avoids aliasing by careful construction of the wavelet filter. Thus, the original signal can be perfectly reconstructed from the downsampled wavelet coefficients. However, any changes made to the wavelet coefficients prior to performing the inverse transform (such as quantization) can cause aliasing in the reconstructed signal. Figure \ref{fig:aliasing} shows the effects of quantization of a simple 1D signal. In the first case, quantization is performed in the sample domain, leading to step edges in the signal. In the second case, a one level wavelet transform is first applied to the signal. Quantization is then performed on the wavelet coefficients prior to reconstruction. This second method of quantization introduces artifacts that did not occur when quantizing in the sample domain.

\section{The Wavelet Transform}

\subsection{1D Wavelet Transform}
The wavelet transform is a linear time-frequency transform that was introduced by Haar \cite{haar1910theorie}. It makes use of a dictionary of functions that are localized in time and frequency. These functions, called wavelets, are dilated and shifted versions of a single mother wavelet function. Our work will be restricted to discrete wavelets which are defined as
\begin{equation}
\psi_j[n] = \frac{1}{2^j} \psi \left( \frac{n}{2^j} \right)
\end{equation}
for $n,j \in \mathds{Z}$. We restrict the wavelets to have unit norm and zero mean. Since the wavelet functions are bandpass, we require the introduction of a scaling function, $\phi$, so that we can can cover all frequencies down to zero. The magnitude of the Fourier transform of the scaling function is set so that \cite{mallat}
\begin{equation}
|\hat{\phi}(\omega)|^2 = \int^{+\infty}_1 \frac{|\hat{\psi}(s\omega)|^2}{s} ds.
\end{equation}

Suppose $x$ is a discrete, uniformly sampled signal of length $N$. The wavelet coefficients are computed by convolving each of the wavelet functions with $x$. The discrete wavelet transform is defined as
\begin{equation}
\label{eq:wt}
W x[n, 2^j] = \sum_{m=0}^{N-1} x[m] \psi_j[m-n].
\end{equation}

The discrete wavelet transform is computed by way of an efficient iterative algorithm. The algorithm makes use of only two filters in order to compute all the wavelet coefficients from Equation \ref{eq:wt}. The first filter,
\begin{equation}
h[n] = \left\langle \frac{1}{\sqrt{2}} \phi\left(\frac{t}{2}\right), \phi(t-n) \right\rangle
\label{eq:wavelet-filter-coefficients}
\end{equation}
is called the scaling (lowpass) filter. The second filter,
\begin{equation}
g[n] =\left\langle  \frac{1}{\sqrt{2}} \psi\left(\frac{t}{2}\right), \phi(t-n) \right\rangle
\label{eq:scaling-filter-coefficients}
\end{equation}
is called the wavelet (highpass) filter.

Each iteration of the algorithm computes the following,
\begin{equation}
\label{eq:dwta}
a_{j+1}[p] = \sum^{+\infty}_{n=-\infty} h[n-2p]a_j[n]
\end{equation}
\begin{equation}
\label{eq:dwtd}
d_{j+1}[p] = \sum^{+\infty}_{n=-\infty} g[n-2p]a_j[n]
\end{equation}
with $a_0=x$. The detail coefficients, $d_j$, correspond exactly to the wavelet coefficients in Equation \ref{eq:wt}. The approximation coefficients, $a_j$, are successively blurred versions of the original signal. The coefficients are downsampled by a factor of two after each iteration.

If the wavelet and scaling functions are orthogonal, we can reconstruct the signal according to
\begin{equation}
\begin{split}
a_{j}[p] =& \sum^{+\infty}_{n=-\infty} h[p-2n]a_{j+1}[n]\\
 		 +& \sum^{+\infty}_{n=-\infty} g[p-2n]d_{j+1}[n]
\end{split}
\label{eq:swt}
\end{equation}
The coefficients must be upsampled at each iteration by inserting zeros at even indices. The reconstruction is called the inverse discrete wavelet transform.

\subsection{2D Wavelet Transform}
\begin{figure}[!tb]
\centering
\subfloat[]{\includegraphics[width=.45\textwidth]{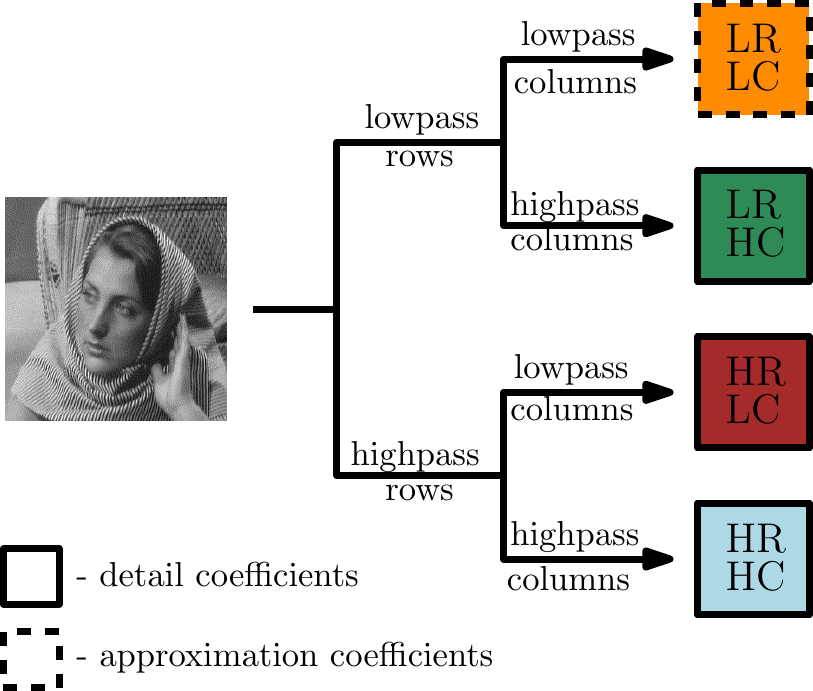}\label{fig:2d-single}}
\hfill
\subfloat[]{\includegraphics[width=.45\textwidth]{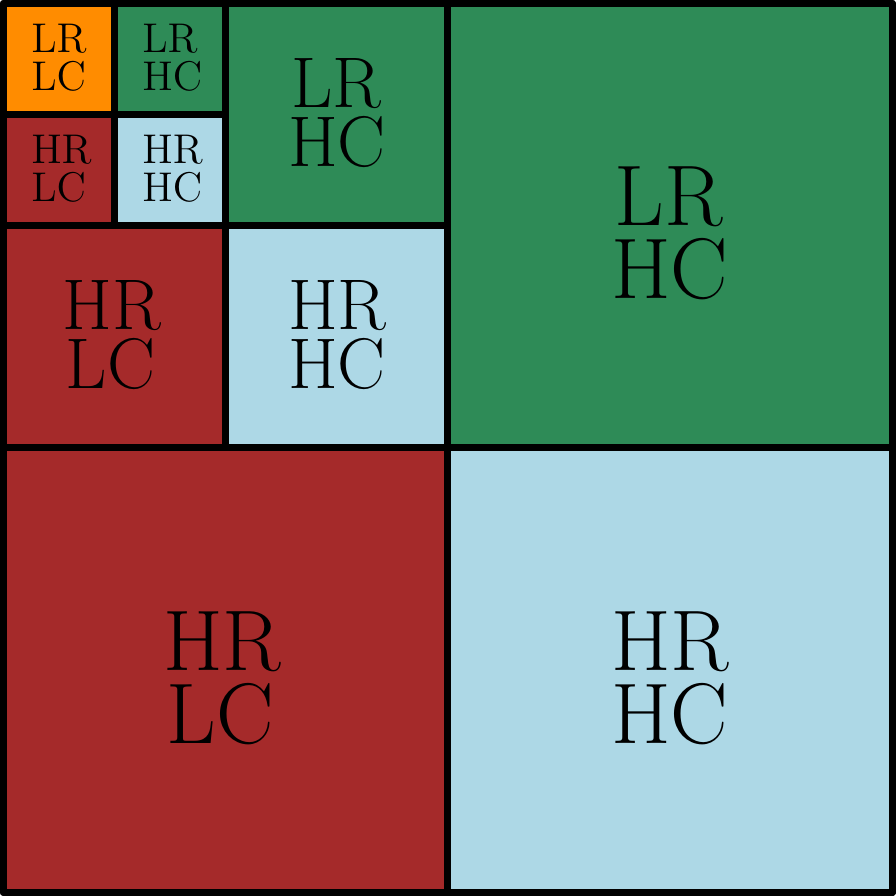}\label{fig:2dtile-a}}
\caption{(a) One iteration of the 2D discrete wavelet transform. (b) Typical arrangement of the wavelet coefficients after three iterations of the algorithm.}
\end{figure}

In order to process 2D signals, such as images, we must use a modified version of the wavelet transform that can be applied to multidimensional signals. The simplest modification is to apply the filters separately along each dimension \cite{mallat1989theory}. Figure \ref{fig:2d-single} shows the coefficients computed for a single iteration of the algorithm, where the input is an image.

Let us define the following four coefficient components
\begin{equation}
\text{approximation: } LR(LC(x))
\label{eq:2d-LL}
\end{equation}
\begin{equation}
\text{detail horizontal: } LR(HC(x))
\label{eq:2d-LH}
\end{equation}
\begin{equation}
\text{detail vertical: } HR(LC(x))
\label{eq:2d-HL}
\end{equation}
\begin{equation}
\text{detail diagonal: } HR(HC(x))
\label{eq:2d-HH}
\end{equation}
where $LR$ and $HR$ correspond to convolving the scaling and wavelet filters respectively along the rows. We can define $LC$ and $HC$ similarly for the columns. At each iteration of the algorithm, we compute the approximation coefficients from Equation \ref{eq:2d-LL} and the three detail coefficient components from Equations \ref{eq:2d-LH}--\ref{eq:2d-HH}. Like in the 1D case, the coefficients are subsampled after each convolution. The coefficients are typically arranged as in Figure \ref{fig:2dtile-a}. Performing the 2D wavelet transform in this manner is used in the JPEG2000 standard  \cite{christopoulos2000jpeg2000}.

\subsection{Dual-Tree Complex Wavelet Transform}
\begin{figure}[!tb]
\centering
\subfloat[]{\includegraphics[width=.3\textwidth]{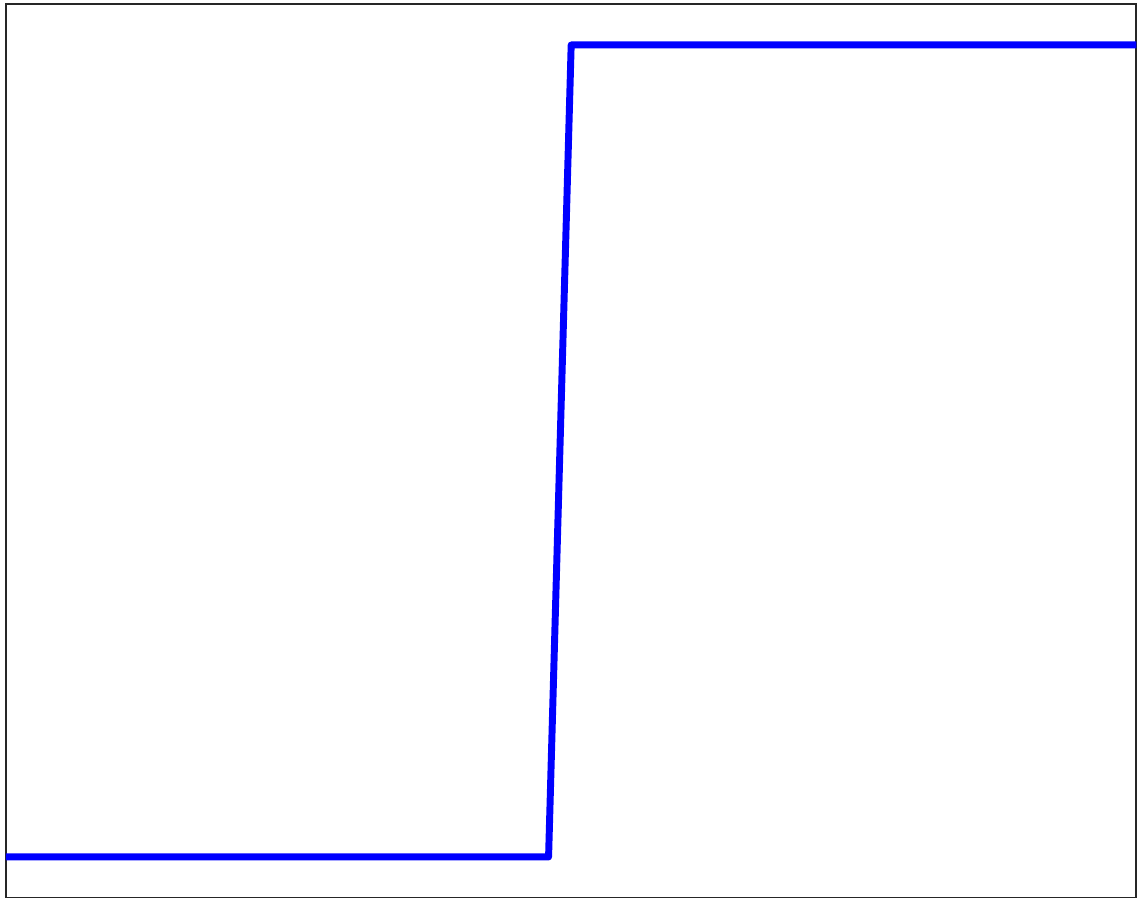}}
\hfill
\subfloat[]{\includegraphics[width=.3\textwidth]{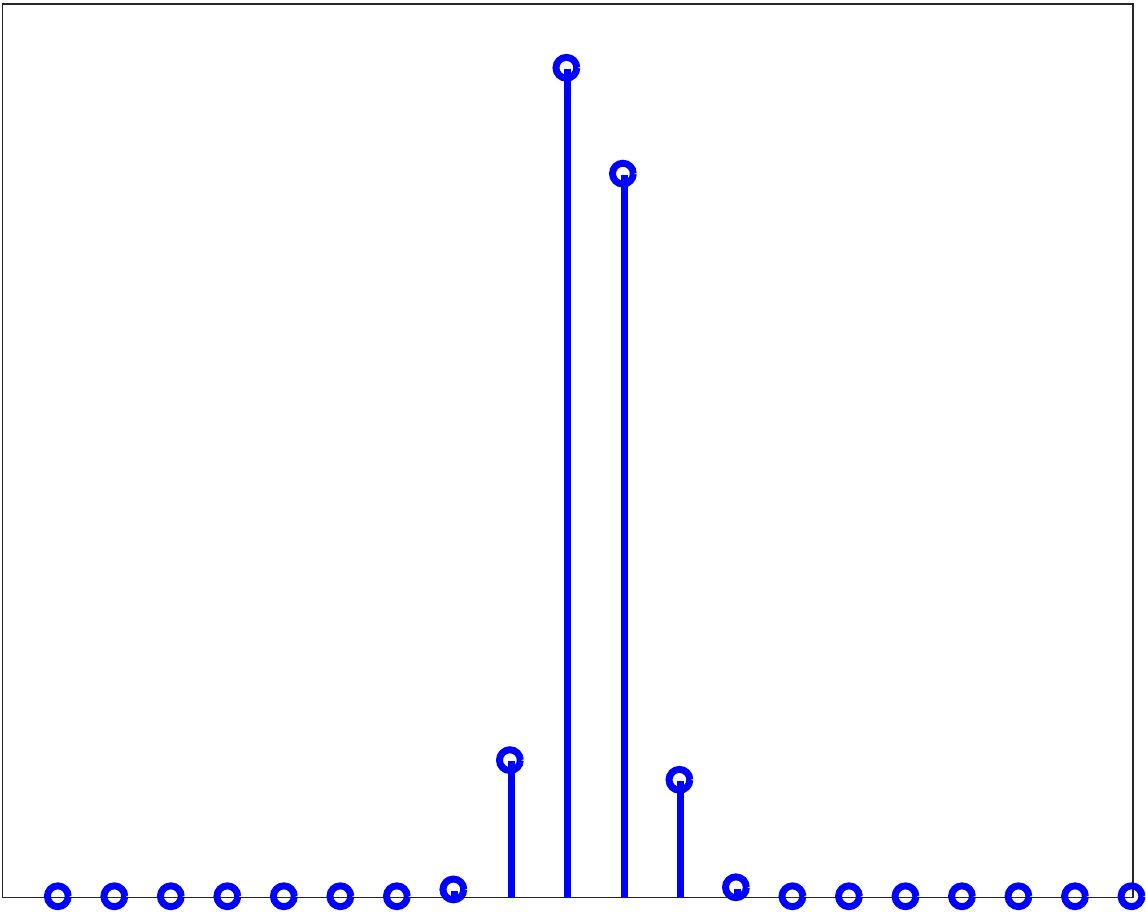}}
\hfill
\subfloat[]{\includegraphics[width=.3\textwidth]{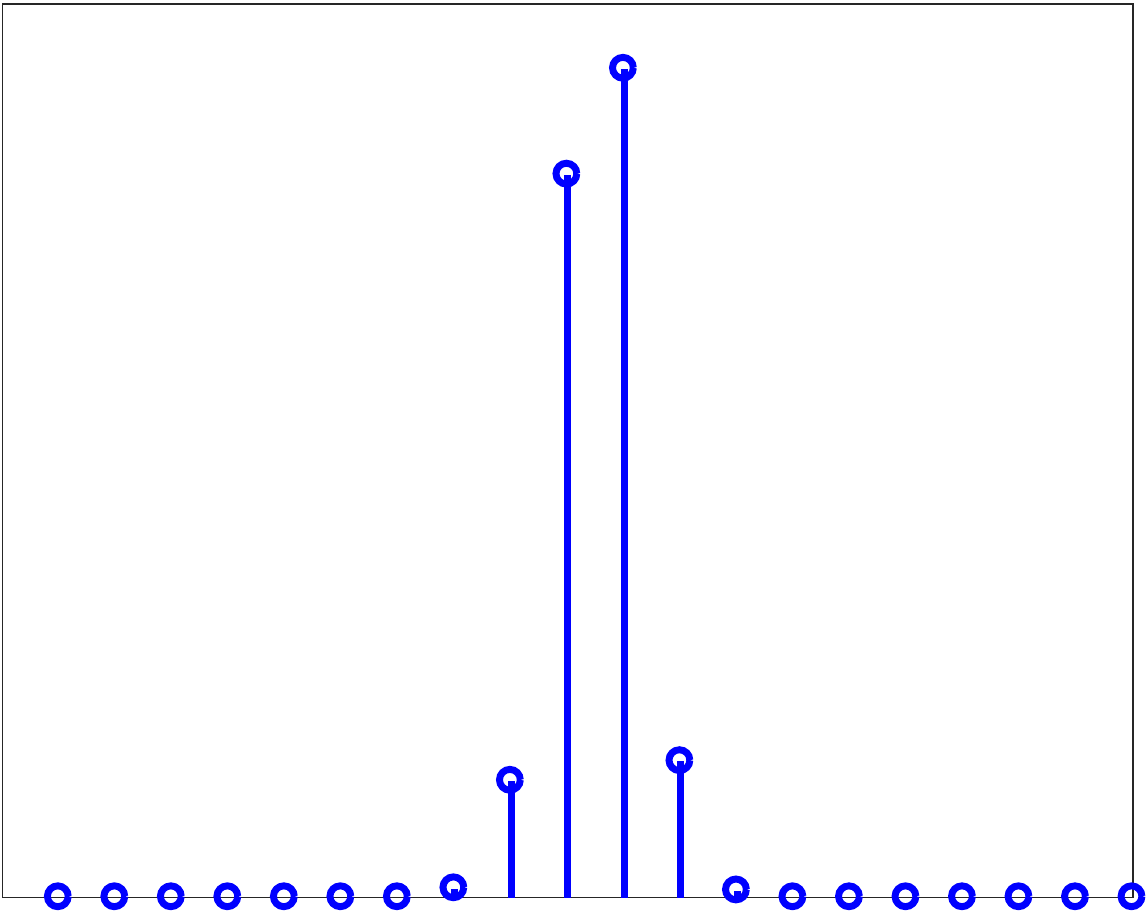}}
\caption{(a) A signal containing a single step edge. (b) The magnitude of the third level DTCWT coefficients of (a). (c) Same as (b) but with the signal in (a) shifted by a single sample.}
\label{fig:qs-shifts}
%
\centering
\subfloat[]{
\includegraphics[width=.22\textwidth,trim={17cm 17cm 0 0},clip]{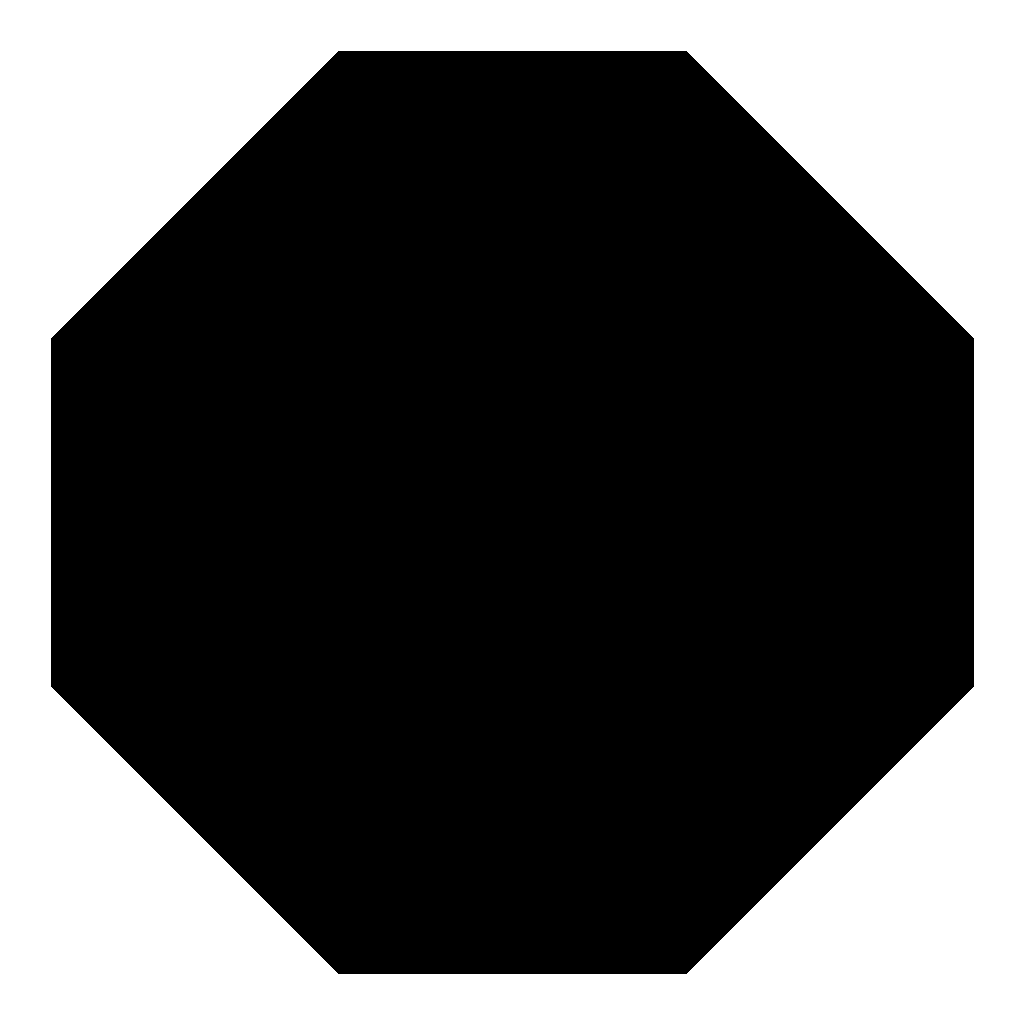}
\hfill
\includegraphics[width=.22\textwidth,trim={17cm 17cm 0 0},clip]{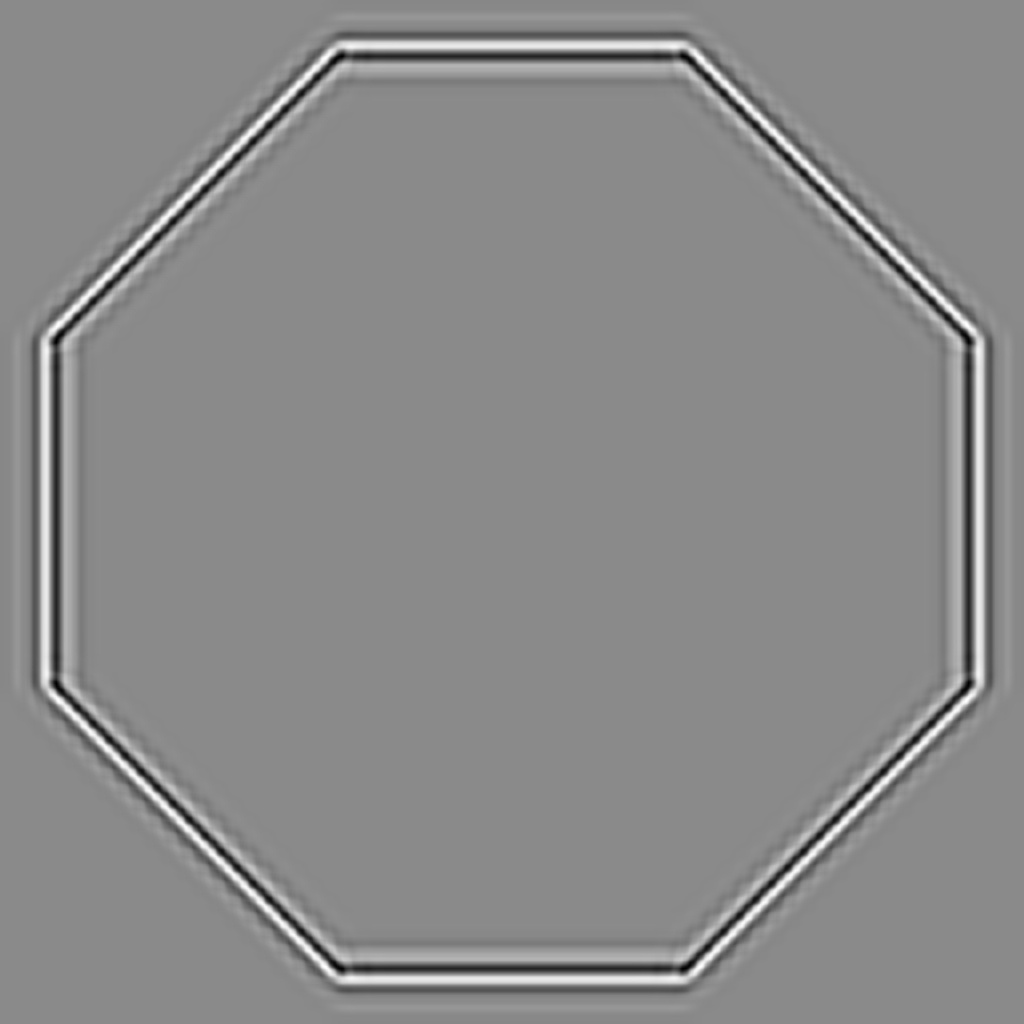}}
\hfill
\subfloat[]{
\includegraphics[width=.22\textwidth,trim={17cm 17cm 0 0},clip]{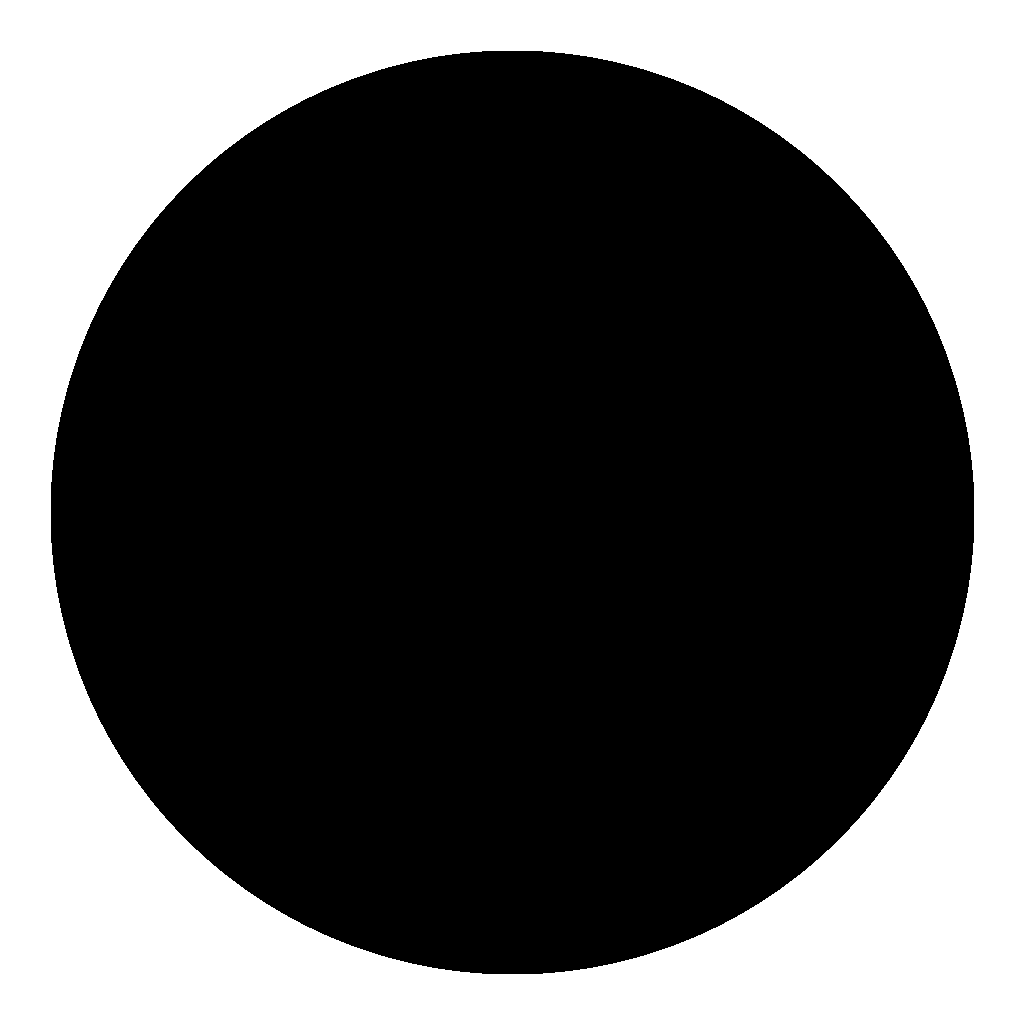}
\hfill
\includegraphics[width=.22\textwidth,trim={17cm 17cm 0 0},clip]{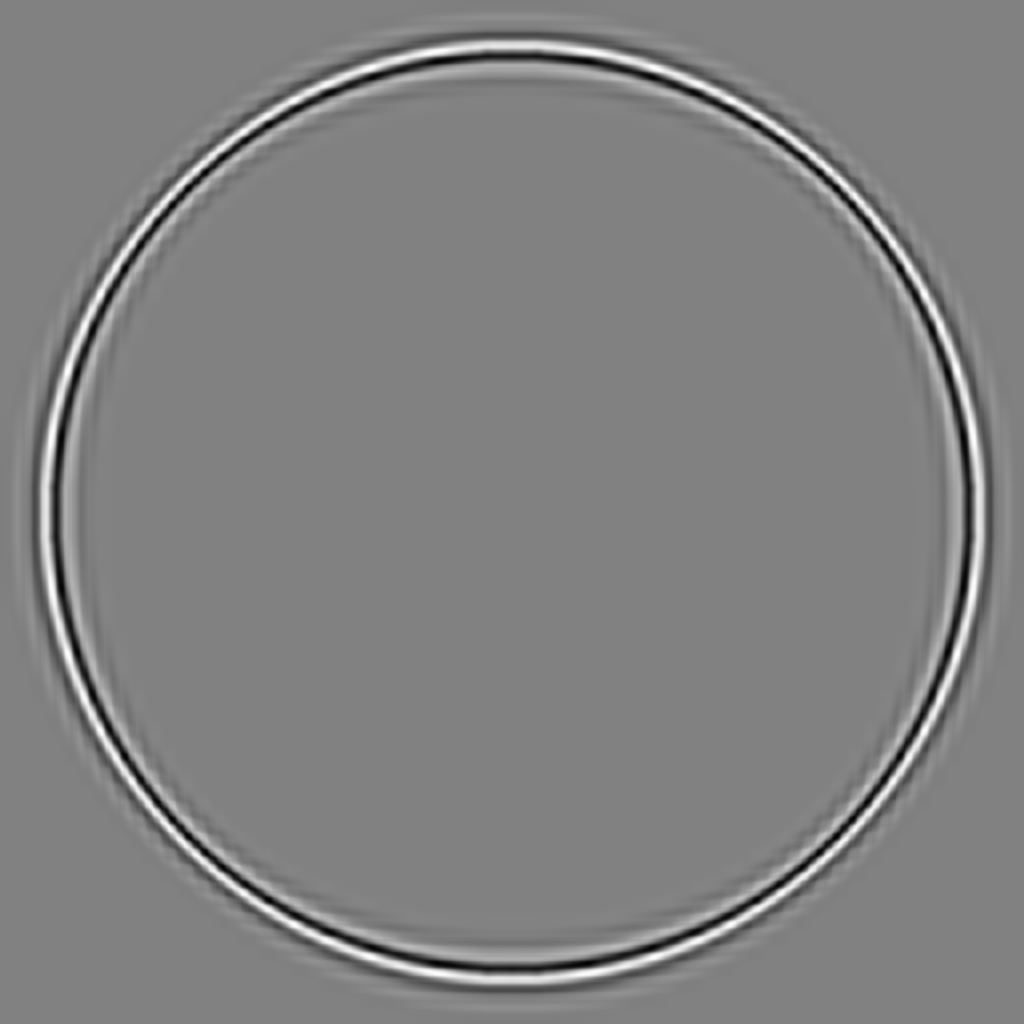}}
\caption{(a) Left: An image with three edge orientations. Right: Reconstruction of the image using only the fourth band DTCWT coefficients.}
\label{fig:qs-bad-edges}
%
\centering
\subfloat[]{\includegraphics[width=.29\textwidth]{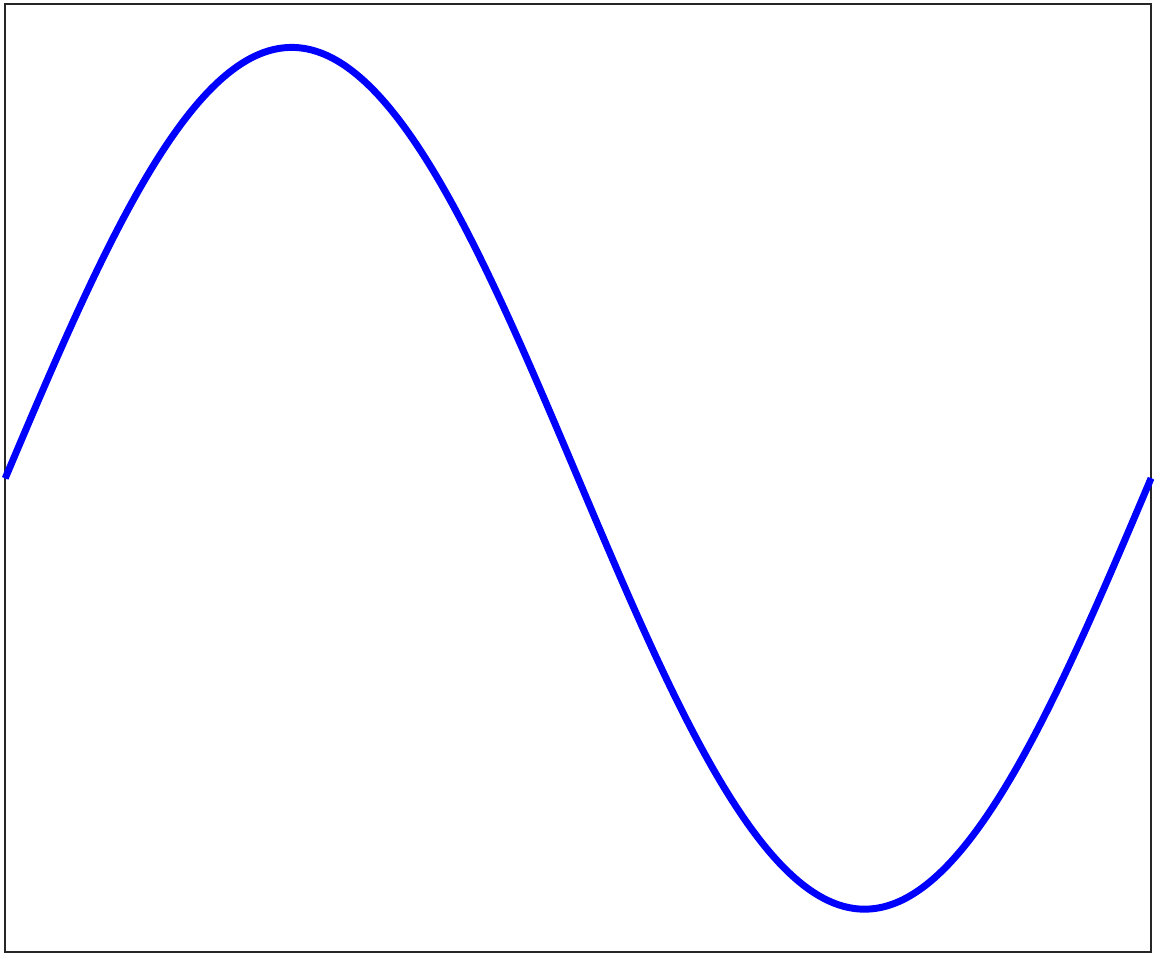}}
\hfill
\subfloat[]{\includegraphics[width=.3\textwidth]{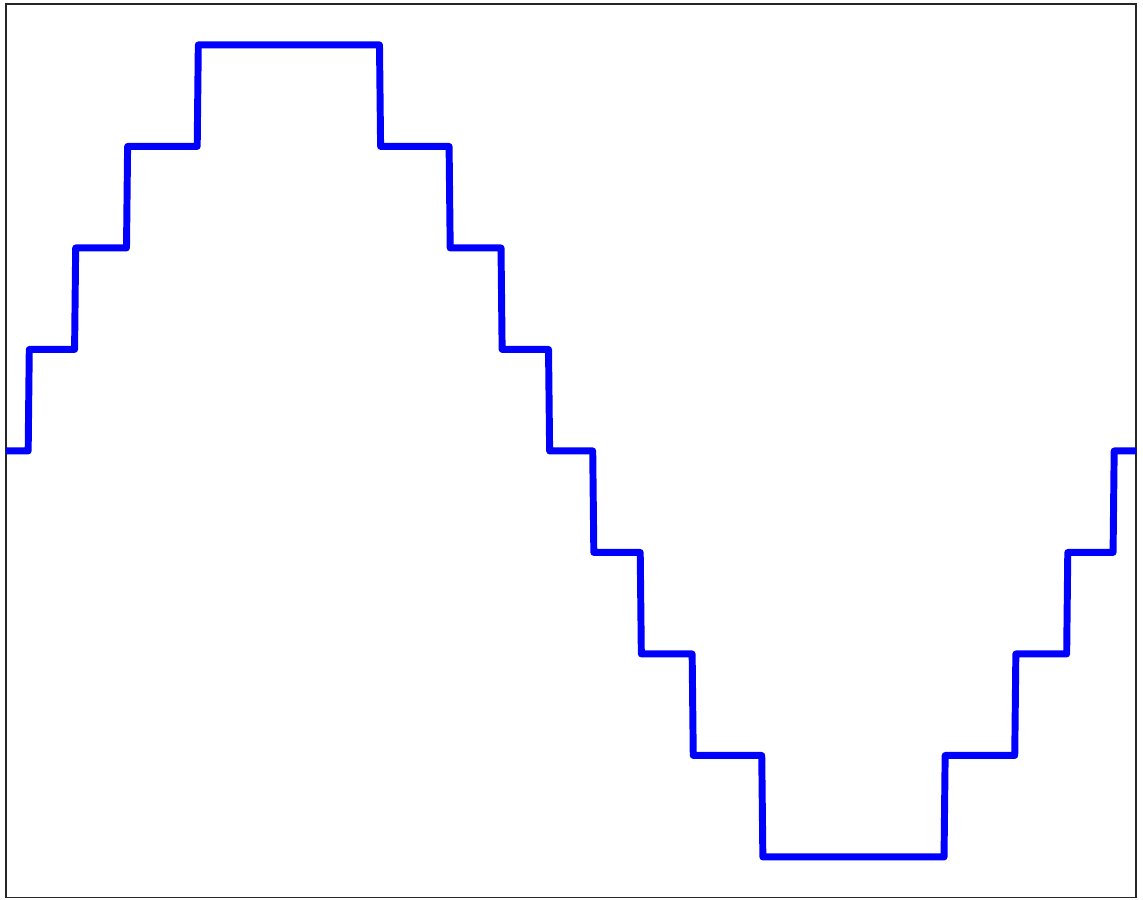}}
\hfill
\subfloat[]{\includegraphics[width=.3\textwidth]{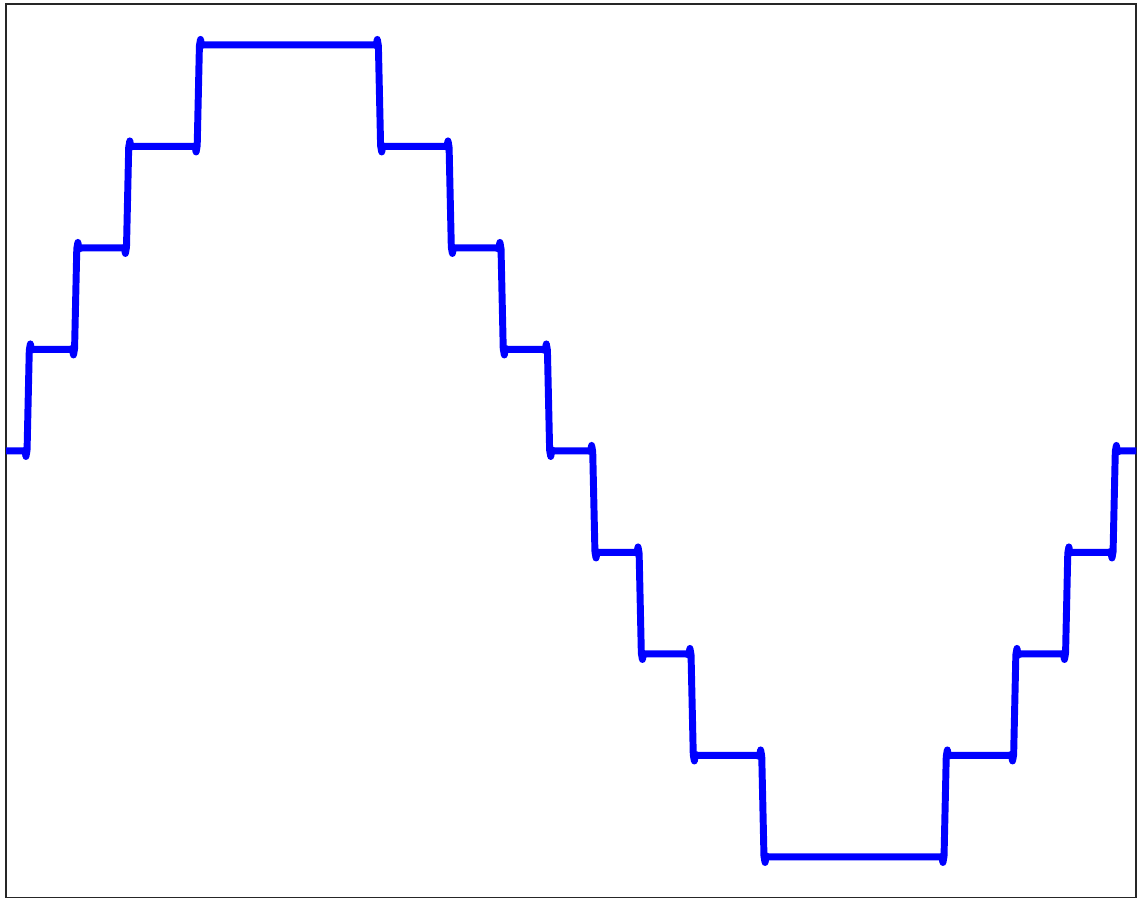}}
\caption{(a) Original signal. (b) The signal from (a) quantized to nine levels. (c) A one level DTCWT is first applied to (a), the wavelet coefficients are quantized to nine levels, and an inverse transform is applied. Note how the artifacts are much less pronounced.}
\label{fig:qs-aliasing}
\end{figure}

The problems with the standard wavelet transform discussed in the previous section can be overcome with the DTCWT \cite{selesnick2005dual} (see Figures \ref{fig:qs-shifts}, \ref{fig:qs-bad-edges}, and \ref{fig:qs-aliasing}). We will restrict our discussion to the 2D version of the transform. As its name suggests, the dual-tree wavelet transform makes use of multiple computational trees. The trees each use a different wavelet and scaling filter pair.

We will begin our discussion with the real version of the dual-tree transform. Let us denote the filters $h_i$ and $g_i$, where ${i=1}$ for the first tree and ${i=2}$ for the second tree. The dual-tree algorithm proceeds similarly to the 2D version. Like in the 2D case, we compute four components of coefficients for each tree: $LR_i(LC_i(x))$, $LR_i(HC_i(x))$, $HR_i(LC_i(x))$, and $HR_i(HC_i(x))$. Each tree computes its own coefficient matrix $W_i$. The final coefficient matrices are computed as follows:
\begin{equation}
W_1 \leftarrow \frac{W_1(x)+W_2(x)}{\sqrt{2}}
\label{eq:wsum}
\end{equation}
\begin{equation}
W_2 \leftarrow \frac{W_1(x)-W_2(x)}{\sqrt{2}}
\label{eq:wdiff}
\end{equation}
We therefore have six bands of detail coefficients at each level of the transform, as opposed to three in the 2D case. Example impulse responses for real filters is shown in Figure \ref{fig:learned-impulses-good-a}. Note that they are oriented along six different directions and there is no checkerboard effect. The drawback of the real dual-tree transform is that it is not approximately shift invariant \cite{selesnick2005dual}.

The complex transform is similarly computed. The main difference is that there are a total of four trees instead of two (for a total of twelve detail bands). The twelve bands are of the form
\begin{equation}
LR_i(HC_j(x)), HR_i(LC_j(x)), HR_i(HC_j(x))
\end{equation} 
for $i,j \in \{1,2\}$. Let $W_{i,j}$ represent the three bands of detail coefficients using filter $i$ for the rows and filter $j$ for the columns. The final complex detail coefficients are computed as:
\begin{equation}
W_{1,1} \leftarrow \frac{W_{1,1}(x)+W_{2,2}(x)}{\sqrt{2}}
\label{eq:wsum1}
\end{equation}
\begin{equation}
W_{2,2} \leftarrow \frac{W_{1,1}(x)-W_{2,2}(x)}{\sqrt{2}}
\label{eq:wdiff1}
\end{equation}
\begin{equation}
W_{1,2} \leftarrow \frac{W_{1,2}(x)+W_{2,1}(x)}{\sqrt{2}}
\label{eq:wsum2}
\end{equation}
\begin{equation}
W_{2,1} \leftarrow \frac{W_{1,2}(x)-W_{2,1}(x)}{\sqrt{2}}
\label{eq:wdiff2}
\end{equation}

In this work we will focus on $q$-shift filters \cite{selesnick2005dual}. That is, we restrict
\begin{equation}
h_2[n] = h_1[-n].
\label{eq:qshift}
\end{equation}
In other words, the filters used in the second tree are the reverse of the filters used in the first tree. Furthermore, the filters used in the first iteration of the algorithm must be different than the filters used for the remaining iterations \cite{selesnick2005dual}. The initial filters must also obey the $q$-shift property, but be offset by one sample. We will denote these filters using a prime symbol (e.g., $h'_i$ is the initial filter in the $i^{th}$ tree). Finally, we restrict ourselves to quadrature mirror filters, i.e.,
\begin{equation}
g[n] = (-1)^n h[-n].
\label{eq:mirror}
\end{equation}
Thus, we can derive the wavelet filters directly from the scaling filter by reversing it and negating alternating indices. Equations \ref{eq:qshift} and \ref{eq:mirror} mean we can completely define the transform with the filters $h_1$ and $h'_1$.

\section{The Wavelet Transform as a Neural Network}
\begin{figure}[!tb]
\centering
\includegraphics[width=1\textwidth]{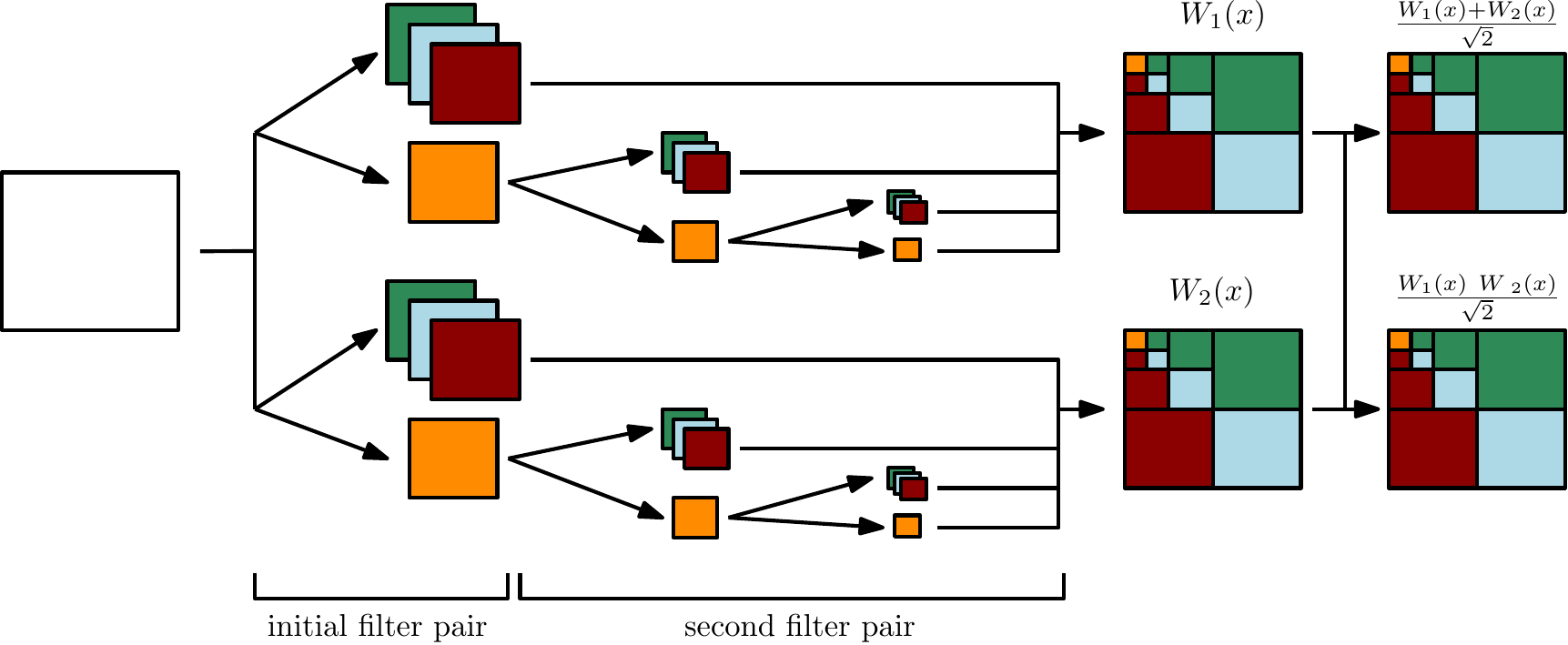}
\caption{The real dual-tree wavelet transform network. Each $W_1$ and $W_2$ are computed as in Equations \ref{eq:wsum} and \ref{eq:wdiff}.}
\label{fig:2dwnn-real}
\end{figure}

\begin{figure}[!tb]
\centering
\includegraphics[width=1\textwidth]{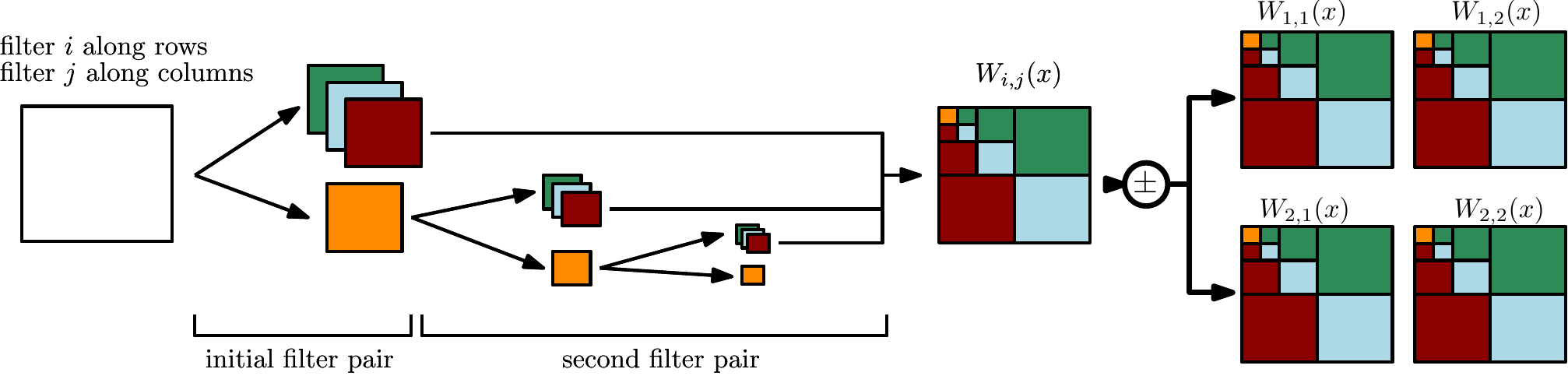}
\caption{The complex dual-tree wavelet transform network. Each $W_{i,j}$ is computed as in Equations \ref{eq:wsum1}--\ref{eq:wdiff2}.}
\label{fig:2dwnn-complex}
\end{figure}

The dual-tree wavelet transform is computed using two main operations: convolution and downsampling. These two operations are the same as those used in traditional convolutional neural networks (CNNs). With this observation in mind, we propose framing the dual-tree wavelet transform as a modified CNN architecture. See Figures \ref{fig:2dwnn-real} and \ref{fig:2dwnn-complex} for illustrations of the real and complex versions of the network respectively. This network directly implements the wavelet transform algorithm, with each layer representing a single iteration. The output of the network are the wavelet coefficients of the input image.

To demonstrate how this model behaves, we construct an autoencoder framework \cite{hinton2006reducing} consisting of a dual-tree wavelet network followed by an inverse network. The intermediate representation in the autoencoder are exactly the wavelet coefficients of the input image. We will impose a sparsity constraint on the wavelet coefficients so that the model will learn filters that can summarize the structure of the training set.

In order for our learned filters to compute a valid wavelet transform, we use the following constraints on the filters \cite{recoskie2018alearning}:
\begin{equation}
L_w(h) = ~(||h||_2 - 1)^2 + (\mu_h - \sqrt{2}/k)^2 + \mu_g^2
\label{eq:wavelet_constraints}
\end{equation}
where $\mu_h$ and $\mu_g$ are the means of the scaling and wavelet filters respectively. The first two terms are necessary so that the scaling function will have unit $L_2$ norm and finite $L_1$ norm \cite{mallat1989multiresolution, mallat}. The third term requires that the wavelet filter has zero mean. 

In order to avoid degenerate filters, we will require an additional loss term. We desire filters that will give localized and centered impulse responses. Therefore, we propose a loss that prefers impulse responses that are close to Gaussian. Let $M^k$ be the matrix corresponding to the magnitude of the impulse response for the $k^{th}$ band (i.e. the sum of the squares of the real and imaginary impulse responses). And let $G$ be a circular Gaussian matrix of the form:
\begin{equation}
G_{i,j} = \alpha e^{-[(i-i_o)^2+(j-j_0)^2]/(2\sigma^2)}
\end{equation}
with $i_0$ and $j_0$ chosen to center the Gaussian in the matrix. The new loss term is of the form
\begin{equation}
L_g = \sum_{k=1}^{6} ||G - M^k||_2^2
\label{eq:gaussian-loss}
\end{equation}

We will learn the network parameters using gradient descent, and so must define a loss function over a dataset of images $X=\{x_1,x_2,\ldots x_M\}$:
\begin{equation}
\begin{split}
L(X; h_1, h'_1) =& \frac{1}{M}\sum_{k=1}^{M} ||x_k - \hat{x}_k||_2^2 \\
             +& \lambda_1 \frac{1}{M}\sum_{k=1}^{M}\sum_{i,j} ||W_{i,j}(x_k)||_1 \\
		+& \lambda_2 [L_w(h_1) + L_w(h'_1)]  + \lambda_3 L_g
\end{split}
\label{eq:loss}
\end{equation}
where $\hat{x}_k$ is the reconstruction of the autoencoder. We use mean squared error for our reconstruction loss and the $L_1$ norm for the sparsity penalty. The $\lambda$ parameters control the trade-off between the loss terms. For the real valued network, the index $j$ is removed from the sparsity summation.

\section{Experiments}
\begin{figure}[!tb]
\centering
\includegraphics[width=.33\textwidth,height=.33\textwidth]{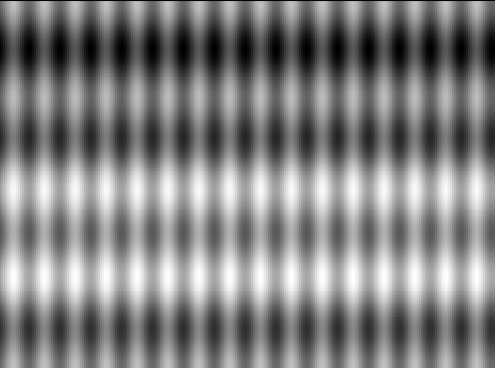}
\hfill
\includegraphics[width=.3\textwidth,height=.33\textwidth]{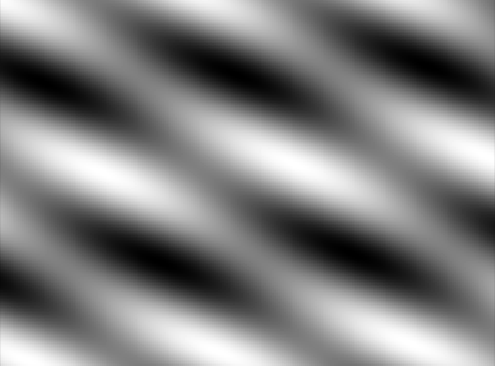}
\hfill
\includegraphics[width=.3\textwidth,height=.33\textwidth]{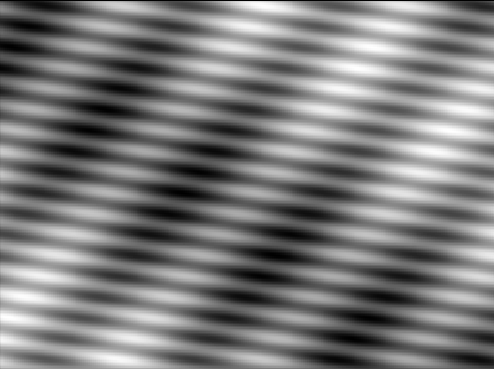}
\caption{Examples of synthetic images.}
\label{fig:data}
\end{figure}

We make use of synthetic images to demonstrate that our model is able to learn filters with localized directional structure. The generation process is the same one used in \cite{recoskie2018blearning}, but with a restriction to sine waves. The process extends the synthetic generation process for 1D data from \cite{recoskie2018alearning}:

\begin{equation}
x(t) = \sum_{k=0}^{K-1} a_k \cdot sin(2^k t + \phi_k)
\label{eq:synth-harmonic}
\end{equation}
where $\phi_k \in [0, 2\pi]$ and $a_k \in \{0, 1\}$ are chosen uniformly at random. In Equation \ref{eq:synth-harmonic}, $\phi_k$ is a phase offset and $a_k$ is an indicator variable that determines whether a particular harmonic is present. We fix the size of the images to be $128 \times 128$. Figure \ref{fig:data} shows some example synthetic images.

We set $\lambda_1=.1$, $\lambda_2=1$, $\lambda_3=4\text{e-5}$ (real network), and $\lambda_3=4\text{e-4}$ (complex network). For the $L_g$ cost, we compute the impulse response at the fourth scale and set $\alpha=.02$ and $\sigma=10$ (found empirically). We train using stochastic gradient descent using the Adam algorithm \cite{kingma2014adam}. Our implementation is done using Tensorflow \cite{tensorflow-short}.

\begin{figure}[!tb]
\centering
\subfloat[Real transform]{\includegraphics[height=.21\textwidth]{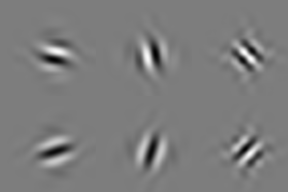}
\label{fig:learned-impulses-good-a}}
\hfill
\subfloat[Complex transform]{\includegraphics[height=.21\textwidth]{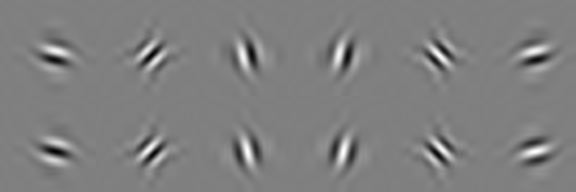}
\label{fig:learned-impulses-good-b}}
\caption{Example impulse responses of filters learned using our (a) real and (b) complex model. The top and bottom rows of (b) correspond to the real and complex parts of the filter responses respectively.}
\label{fig:learned-impulses-good}
\end{figure}

\begin{figure}[!tb]
\centering
\subfloat[]{\includegraphics[width=.5\textwidth]{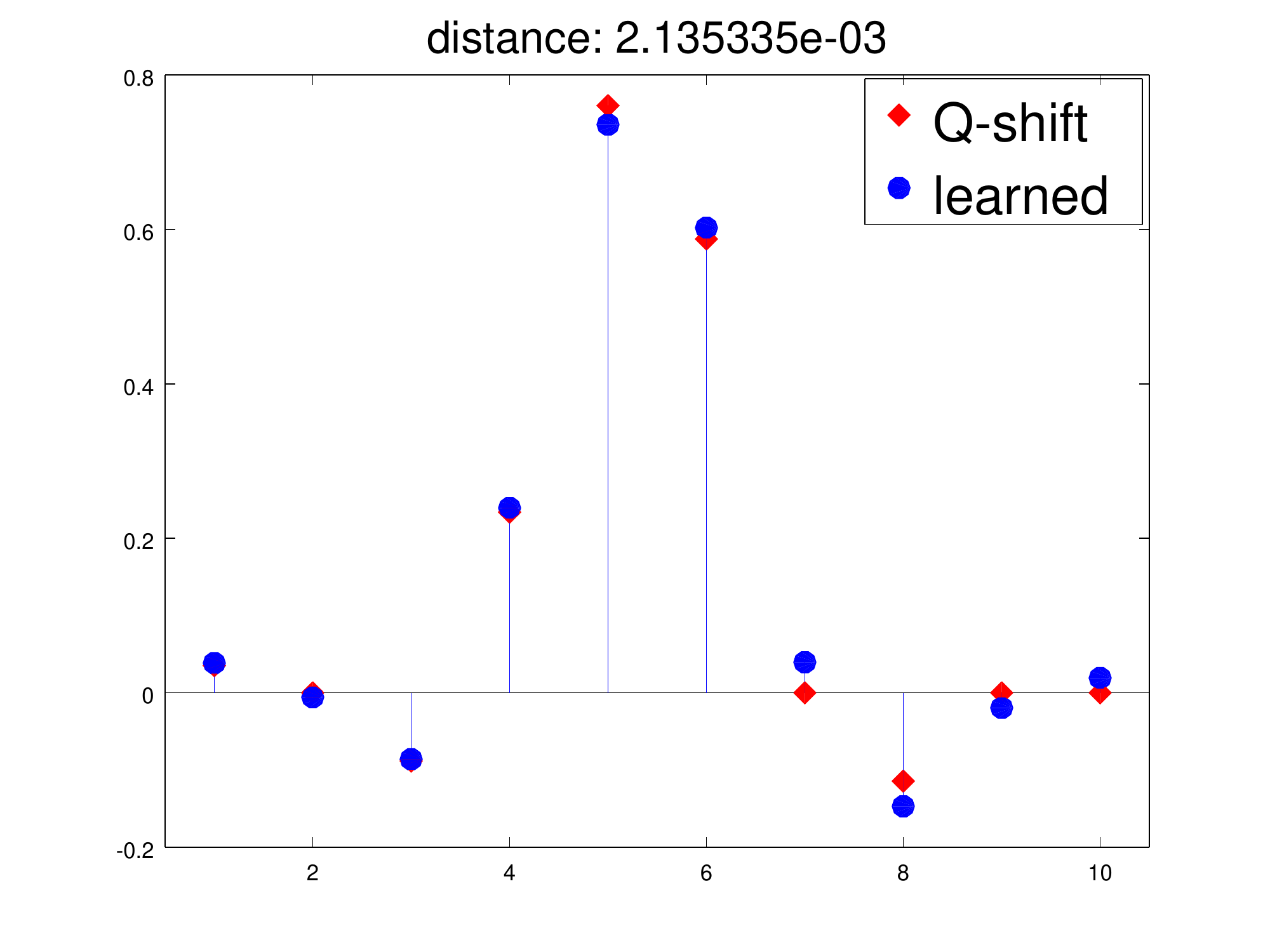}}
\hfill
\subfloat[]{\includegraphics[width=.5\textwidth]{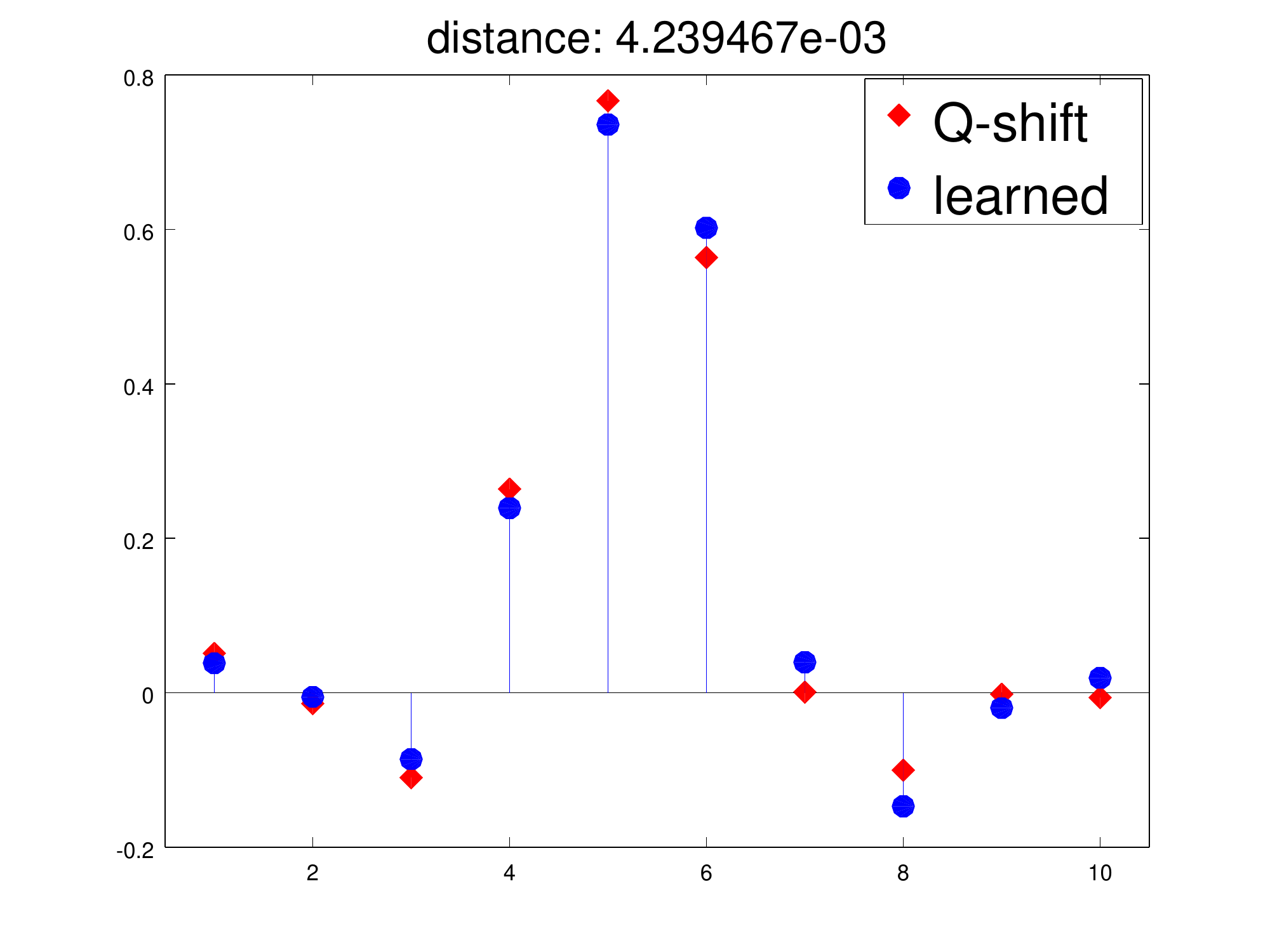}}
\caption{Comparison of Kingsbury's Q-shift (a) 6 tap, and (b) 10 tap filters with the learned $h$ filter from Figure \ref{fig:learned-impulses-good-b}. Note that $h$ has been reversed.}
\label{fig:learned-complex10-main}
\end{figure}

Figure \ref{fig:learned-impulses-good} shows the impulse responses of the filters learned by our real and complex models. In this case, all filters were of length ten. We can see that the model is able to learn localized, directional filters from the data. Figure \ref{fig:learned-complex10-main} shows a comparison of the learned $h$ filter from Figure \ref{fig:learned-impulses-good-b} to Kingsbury's Q-shift filters. Note that $h$ has a very similar structure. The shifted cosine distance measure used in \cite{recoskie2018alearning} is shown above each plot. See the appendix for more filters and plots.

\begin{figure}[!tb]
\centering
\subfloat[]{\includegraphics[height=.21\textwidth]{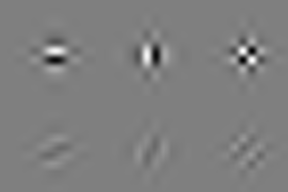}
\label{fig:learned-impulses-degen-a}}
\hfill
\subfloat[]{\includegraphics[height=.21\textwidth]{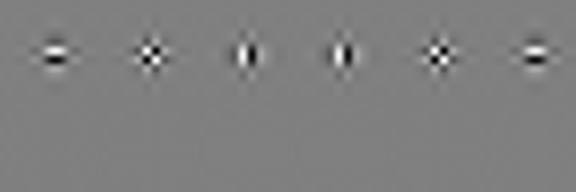}
\label{fig:learned-impulses-degen-b}}\\
\subfloat[]{\includegraphics[height=.21\textwidth]{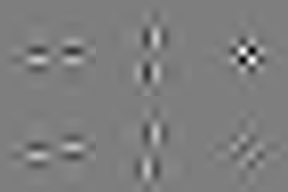}
\label{fig:learned-impulses-degen-c}}
\hfill
\subfloat[]{\includegraphics[height=.21\textwidth]{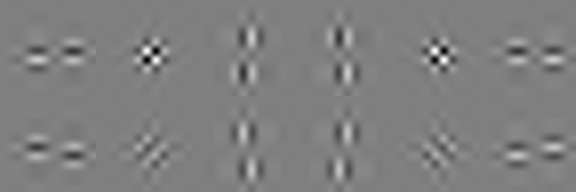}
\label{fig:learned-impulses-degen-d}}\\
\caption{Example impulse responses of filters learned using our (a,c) real and (b,d) complex model without the impulse response loss term (i.e. $\lambda_3 = 0$). }
\label{fig:learned-impulses-degen}
\end{figure}

The Gaussian loss term is important for learning directional filters. Figure \ref{fig:learned-impulses-degen} show a selection of filters learned without the impulse response loss term (i.e. $\lambda_3=0$ in Equation \ref{eq:loss}). We can see that the filters are no longer directional, and share similar structure to the impulse responses from Figure \ref{fig:oriented}.

\section{Conclusion}
We presented a model based on the dual-tree complex wavelet transform that is capable of learning directional filters that overcome the limitations of the standard 2D wavelet transform. We made use of an autoencoder framework, and constrained our loss function to favour orthogonal wavelet filters that produced sparse representations. This work builds upon our previous work on the 1D and 2D wavelet transform. We propose this method as an alternative to the traditional filter design methods used in wavelet theory.

\clearpage
\begin{appendices}
\section{Learned Filter Values}
We include the learned filter values from Figures \ref{fig:learned-impulses-degen} and \ref{fig:learned-impulses-good}.

\subsection{Real dual-tree filters}
\begin{table}[H]
\noindent\makebox[\textwidth]{%
\captionsetup[subfloat]{labelformat=empty}
\centering
\subfloat[Filters from Figure \ref{fig:learned-impulses-good-a} 2.]{
\begin{tabular}{ | c | c | }
\hline
$h'_1$ & $h_1$ \\
\hline
$+7.538\text{e-04}$ & $+3.207\text{e-02}$ \\
$-6.960\text{e-02}$ & $-4.052\text{e-03}$ \\
$-4.274\text{e-02}$ & $-5.705\text{e-02}$ \\
$+4.233\text{e-01}$ & $+2.732\text{e-01}$ \\
$+7.917\text{e-01}$ & $+7.357\text{e-01}$ \\
$+4.234\text{e-01}$ & $+5.615\text{e-01}$ \\
$-4.291\text{e-02}$ & $+5.348\text{e-03}$ \\
$-6.972\text{e-02}$ & $-1.456\text{e-01}$ \\
$+2.573\text{e-04}$ & $-5.864\text{e-03}$ \\
$+4.025\text{e-04}$ & $+2.406\text{e-02}$ \\
\hline
\end{tabular}
}
\hfill
\subfloat[Filters from Figure \ref{fig:learned-impulses-degen-a}.]{
\begin{tabular}{ | c | c | }
\hline
$h'_1$ & $h_1$ \\
\hline
$-1.010\text{e-02}$ & $+1.970\text{e-02}$ \\
$+3.534\text{e-02}$ & $-3.822\text{e-02}$ \\
$+2.407\text{e-02}$ & $-1.078\text{e-01}$ \\
$-8.123\text{e-02}$ & $+2.241\text{e-01}$ \\
$+2.722\text{e-01}$ & $+7.332\text{e-01}$ \\
$+7.981\text{e-01}$ & $+6.144\text{e-01}$ \\
$+5.143\text{e-01}$ & $+5.249\text{e-02}$ \\
$-4.624\text{e-02}$ & $-1.302\text{e-01}$ \\
$-9.701\text{e-02}$ & $+9.322\text{e-03}$ \\
$-5.738\text{e-04}$ & $+3.693\text{e-02}$ \\
\hline
\end{tabular}
}
\hfill
\subfloat[Filters from Figure \ref{fig:learned-impulses-degen-c}.]{
\begin{tabular}{ | c | c | }
\hline
$h'_1$ & $h_1$ \\
\hline
$+1.956\text{e-02}$ & $+1.970\text{e-02}$ \\
$-5.031\text{e-02}$ & $-3.822\text{e-02}$ \\
$-7.082\text{e-02}$ & $-1.078\text{e-01}$ \\
$+4.035\text{e-01}$ & $+2.241\text{e-01}$ \\
$+8.096\text{e-01}$ & $+7.332\text{e-01}$ \\
$+4.028\text{e-01}$ & $+6.144\text{e-01}$ \\
$-7.107\text{e-02}$ & $+5.249\text{e-02}$ \\
$-4.922\text{e-02}$ & $-1.302\text{e-01}$ \\
$+1.958\text{e-02}$ & $+9.322\text{e-03}$ \\
$-1.094\text{e-05}$ & $+3.693\text{e-02}$ \\
\hline
\end{tabular}
}
}
\end{table}

\begin{figure}[H]
\noindent\makebox[\textwidth]{%
\centering
\subfloat[]{\includegraphics[width=.6\textwidth]{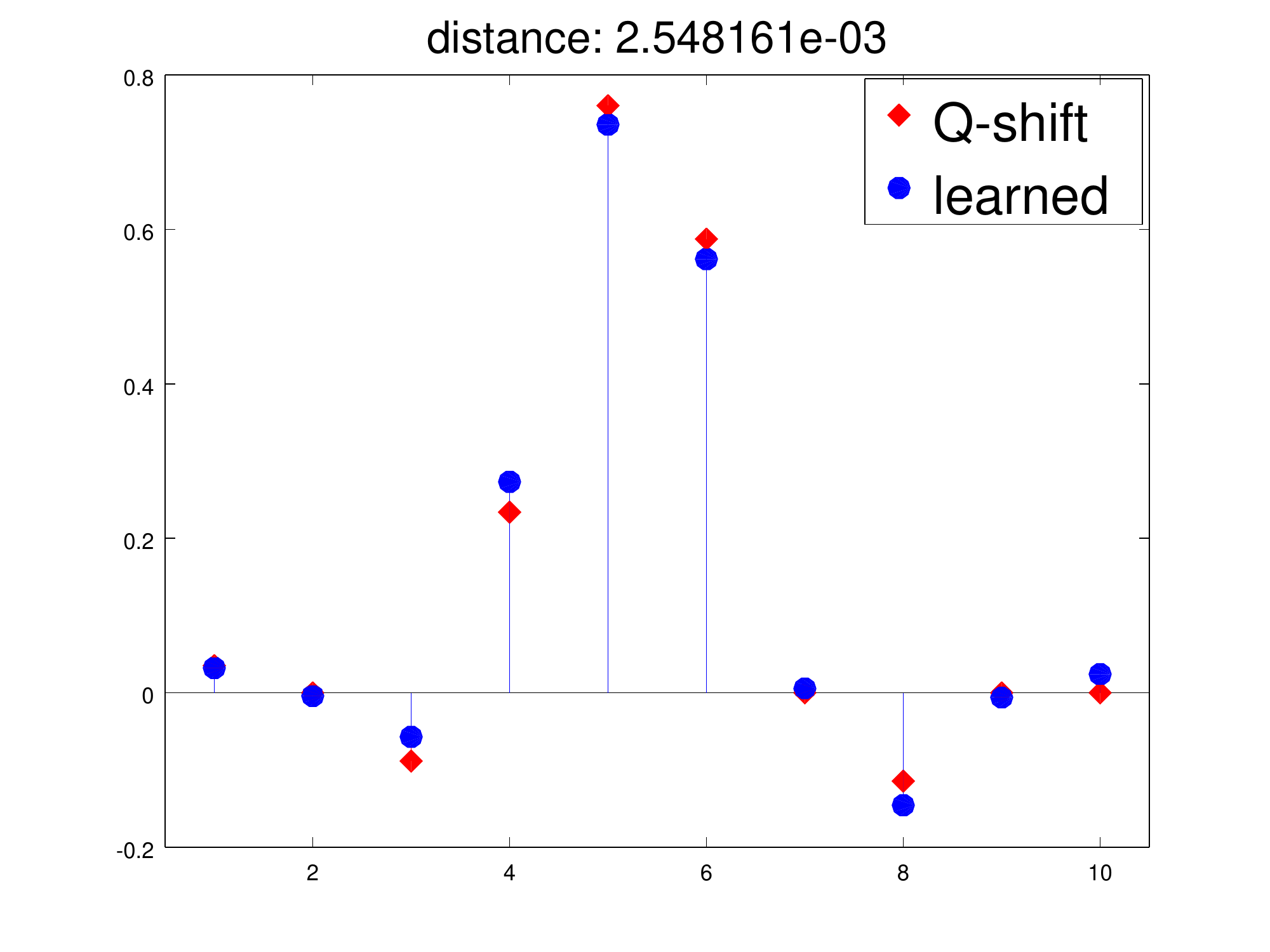}}
\subfloat[]{\includegraphics[width=.6\textwidth]{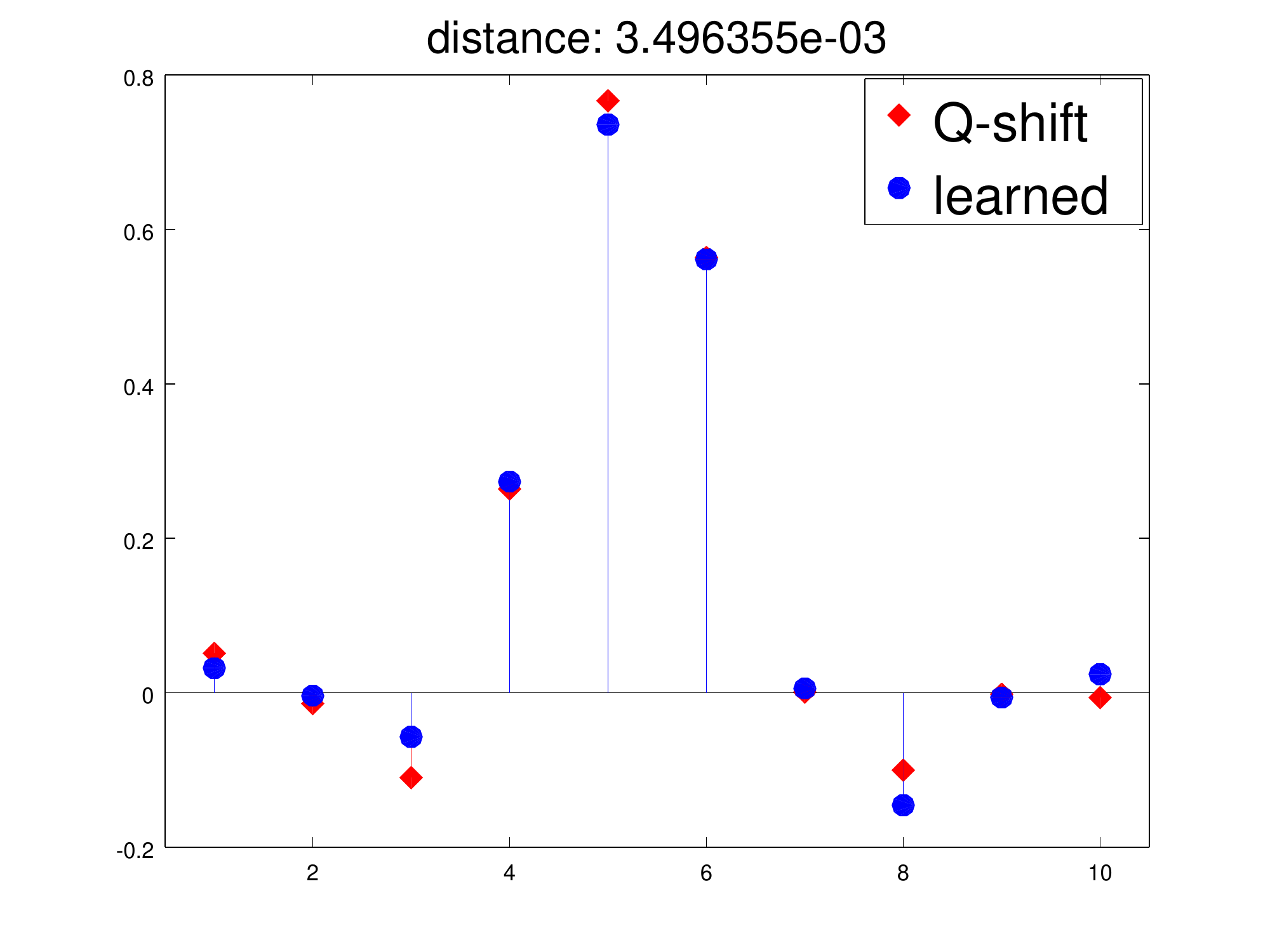}}}
\caption{Comparison of Kingsbury's Q-shift (a) 6 tap, and (b) 10 tap filters with the learned $h$ filter from Figure \ref{fig:learned-impulses-good-a}.}
\label{fig:learned-real10}
\end{figure}

\subsection{Complex dual-tree filters}
\begin{table}[H]
\noindent\makebox[\textwidth]{%
\captionsetup[subfloat]{labelformat=empty}
\centering
\subfloat[Filters from Figure \ref{fig:learned-impulses-good-b}.]{
\begin{tabular}{ | c | c | }
\hline
$h'_1$ & $h_1$ \\
\hline
$+1.017\text{e-02}$ & $+1.914\text{e-02}$ \\
$+3.677\text{e-02}$ & $-1.947\text{e-02}$ \\
$-3.689\text{e-02}$ & $-1.469\text{e-01}$ \\
$-5.913\text{e-02}$ & $+3.960\text{e-02}$ \\
$+3.982\text{e-01}$ & $+6.020\text{e-01}$ \\
$+8.018\text{e-01}$ & $+7.357\text{e-01}$ \\
$+4.223\text{e-01}$ & $+2.393\text{e-01}$ \\
$-7.171\text{e-02}$ & $-8.599\text{e-02}$ \\
$-8.616\text{e-02}$ & $-5.614\text{e-03}$ \\
$+4.248\text{e-05}$ & $+3.850\text{e-02}$ \\
\hline
\end{tabular}
}
\hfill
\subfloat[Filters from Figure \ref{fig:learned-impulses-degen-b}.]{
\begin{tabular}{ | c | c | }
\hline
$h'_1$ & $h_1$ \\
\hline
$+1.523\text{e-02}$ & $+2.826\text{e-02}$ \\
$-5.440\text{e-02}$ & $-1.266\text{e-02}$ \\
$-6.360\text{e-02}$ & $-1.225\text{e-01}$ \\
$+4.075\text{e-01}$ & $+1.364\text{e-01}$ \\
$+8.054\text{e-01}$ & $+6.800\text{e-01}$ \\
$+4.075\text{e-01}$ & $+6.807\text{e-01}$ \\
$-6.346\text{e-02}$ & $+1.358\text{e-01}$ \\
$-5.442\text{e-02}$ & $-1.220\text{e-01}$ \\
$+1.547\text{e-02}$ & $-1.340\text{e-02}$ \\
$+1.666\text{e-04}$ & $+2.921\text{e-02}$ \\
\hline
\end{tabular}
}
\hfill
\subfloat[Filters from Figure \ref{fig:learned-impulses-degen-d}.]{
\begin{tabular}{ | c | c | }
\hline
$h'_1$ & $h_1$ \\
\hline
$+1.903\text{e-02}$ & $-8.475\text{e-02}$ \\
$-4.904\text{e-02}$ & $-8.946\text{e-02}$ \\
$-7.057\text{e-02}$ & $+3.345\text{e-01}$ \\
$+4.024\text{e-01}$ & $+7.659\text{e-01}$ \\
$+8.090\text{e-01}$ & $+5.145\text{e-01}$ \\
$+4.020\text{e-01}$ & $-1.701\text{e-02}$ \\
$-7.052\text{e-02}$ & $-9.557\text{e-02}$ \\
$-4.888\text{e-02}$ & $+5.711\text{e-02}$ \\
$+1.931\text{e-02}$ & $+3.971\text{e-02}$ \\
$+5.240\text{e-04}$ & $-9.468\text{e-03}$ \\
\hline
\end{tabular}
}
}
\end{table}

\begin{figure}[H]
\noindent\makebox[\textwidth]{%
\centering
\subfloat[]{\includegraphics[width=.6\textwidth]{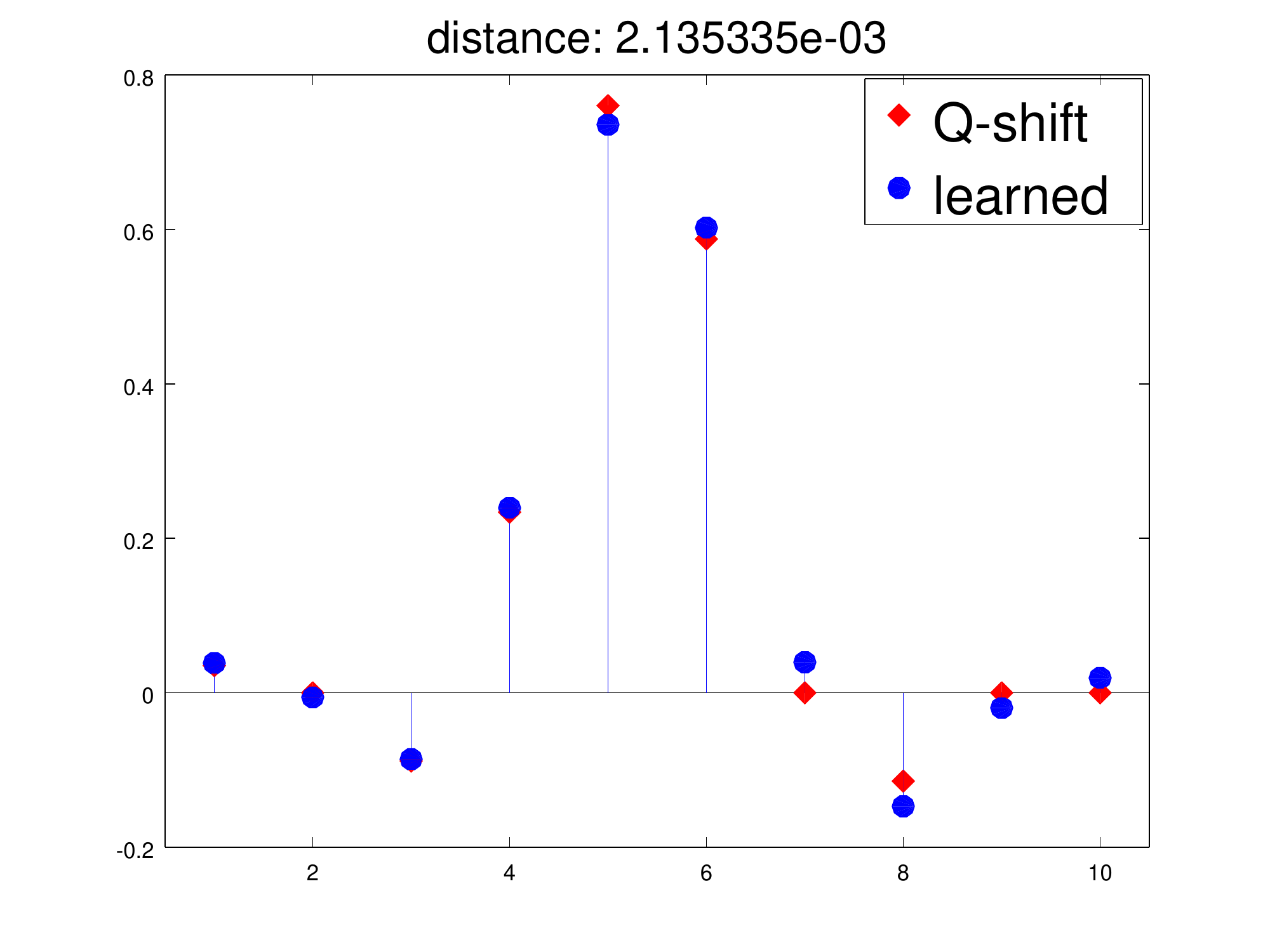}}
\subfloat[]{\includegraphics[width=.6\textwidth]{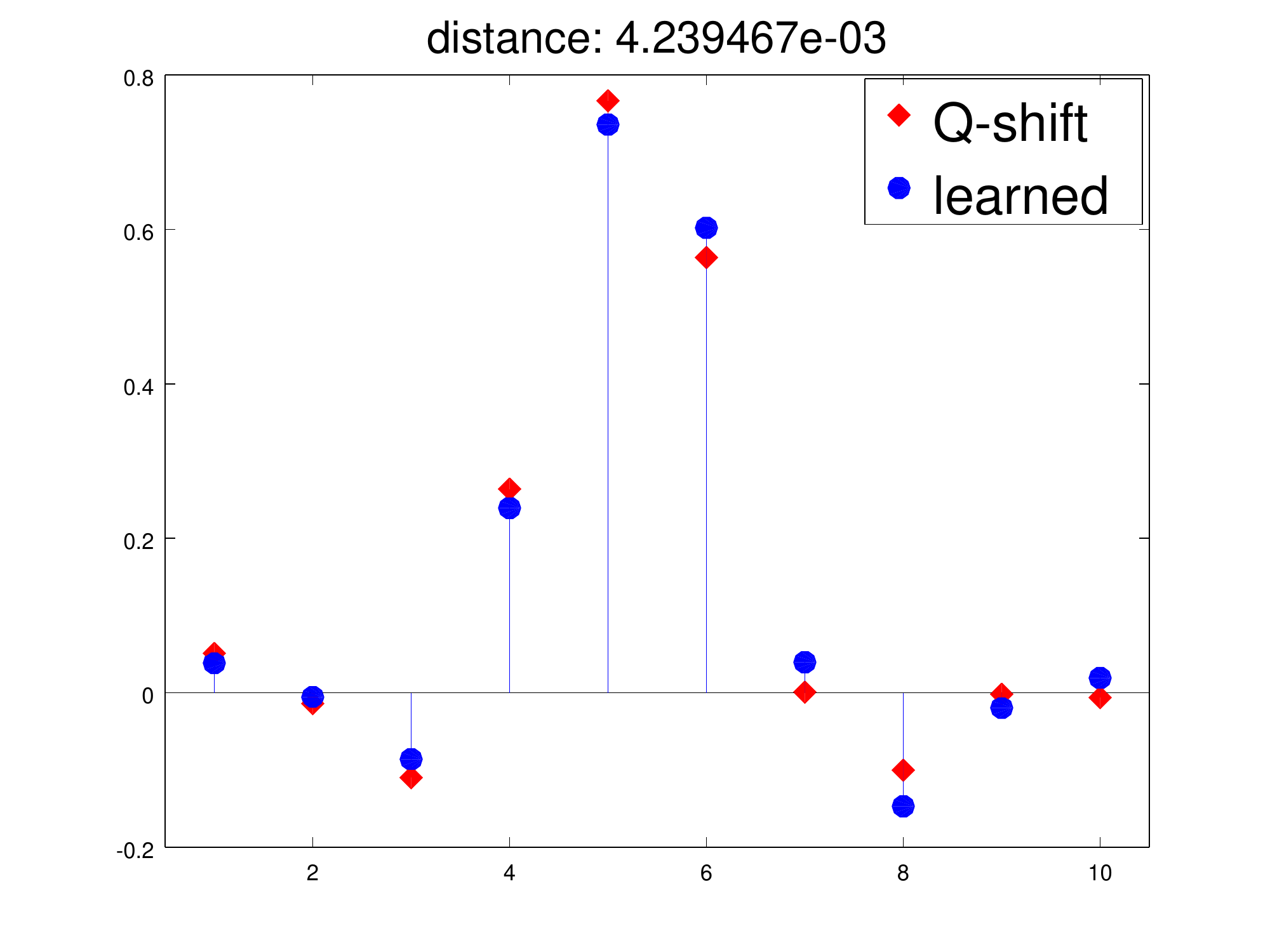}}}
\caption{Comparison of Kingsbury's Q-shift (a) 6 tap, and (b) 10 tap filters with the learned $h$ filter from Figure \ref{fig:learned-impulses-good-b}. Note that $h$ has been reversed.}
\label{fig:learned-complex10}
\end{figure}

\begin{figure}[H]
\centering
\begin{minipage}[c]{0.3\textwidth}
\begin{table}[H]
\begin{tabular}{ | c | c | }
\hline
$h'_1$ & $h_1$ \\
\hline
$+1.640\text{e-02}$ & $-1.571\text{e-03}$ \\
$-5.179\text{e-02}$ & $-2.007\text{e-03}$ \\
$-6.775\text{e-02}$ & $+2.725\text{e-02}$ \\
$+4.049\text{e-01}$ & $-2.220\text{e-03}$ \\
$+8.068\text{e-01}$ & $-1.437\text{e-01}$ \\
$+4.050\text{e-01}$ & $+2.654\text{e-03}$ \\
$-1.189\text{e-01}$ & $+5.598\text{e-01}$ \\
$+1.654\text{e-02}$ & $+7.526\text{e-01}$ \\
$+3.910\text{e-04}$ & $+2.865\text{e-01}$ \\
  & $-7.620\text{e-02}$ \\
  & $-2.374\text{e-02}$ \\
  & $+3.958\text{e-02}$ \\
  & $+2.277\text{e-03}$ \\
  & $-7.738\text{e-03}$ \\
\hline
\end{tabular}
\caption*{Learned length 10 $h'$ and length 14 $h$ filters.}
\end{table}
\end{minipage}
\hfill
\begin{minipage}[c]{0.6\textwidth}
\includegraphics[width=1\textwidth]{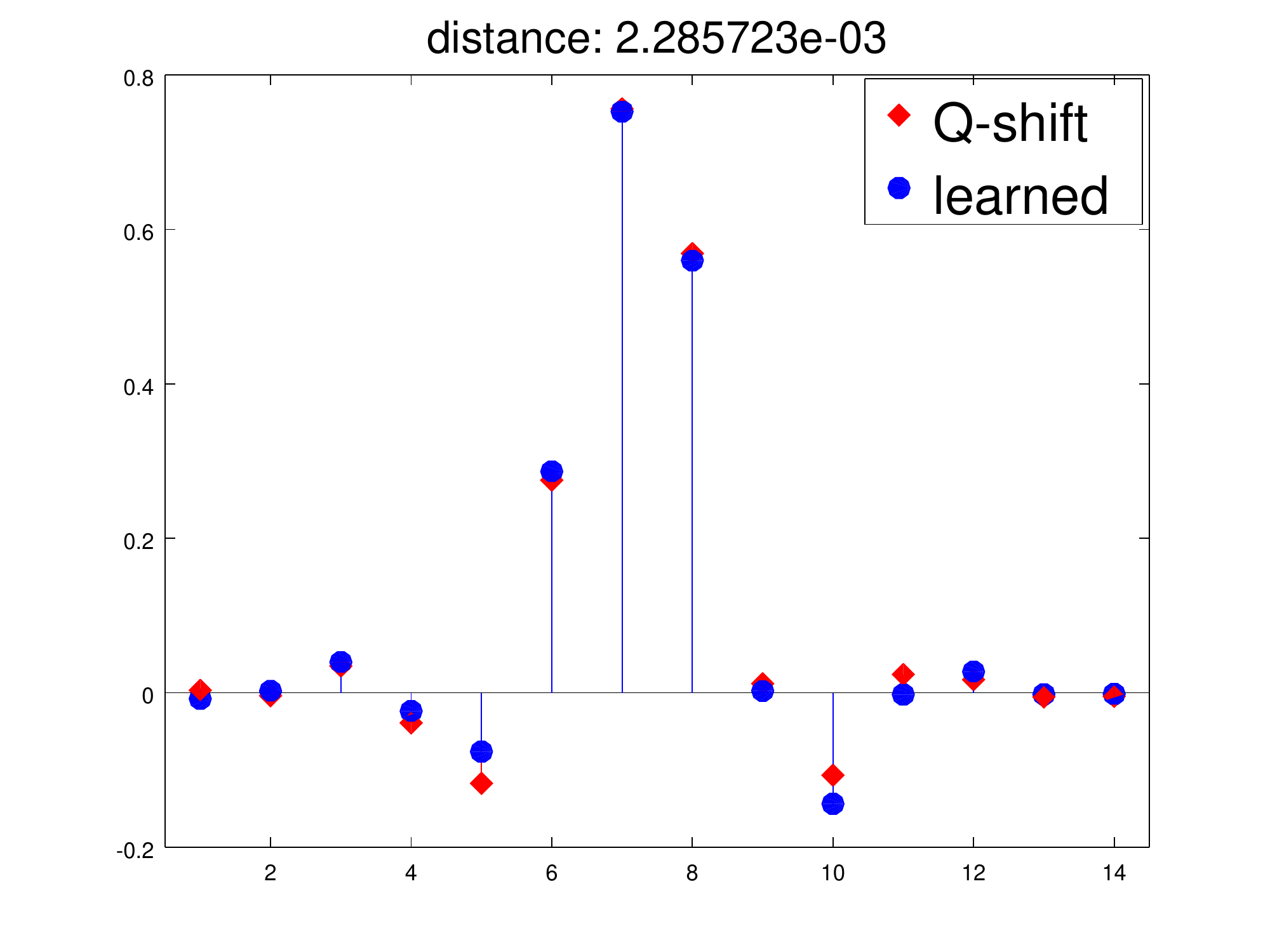}
\caption{Comparison of Kingsbury's Q-shift 14 tap filter with the learned length 14 $h$. Note that $h$ has been reversed.}
\end{minipage}
\end{figure}

\begin{figure}[H]
\centering
\begin{minipage}[c]{0.3\textwidth}
\begin{table}[H]
\begin{tabular}{ | c | c | }
\hline
$h'_1$ & $h_1$ \\
\hline
$+1.952\text{e-02}$ & $+4.335\text{e-04}$ \\
$-4.506\text{e-02}$ & $+3.752\text{e-04}$ \\
$-6.062\text{e-02}$ & $-6.027\text{e-03}$ \\
$+4.069\text{e-01}$ & $+3.851\text{e-03}$ \\
$+8.066\text{e-01}$ & $+3.955\text{e-02}$ \\
$+4.043\text{e-01}$ & $-2.870\text{e-02}$ \\
$-7.323\text{e-02}$ & $-8.351\text{e-02}$ \\
$-6.027\text{e-02}$ & $+2.887\text{e-01}$ \\
$+1.452\text{e-02}$ & $+7.560\text{e-01}$ \\
$-9.868\text{e-05}$ & $+5.569\text{e-01}$ \\
  & $+3.984\text{e-03}$ \\
  & $-1.344\text{e-01}$ \\
  & $-1.460\text{e-03}$ \\
  & $+2.056\text{e-02}$ \\
  & $-2.688\text{e-03}$ \\
  & $-5.462\text{e-04}$ \\
  & $-1.798\text{e-03}$ \\
  & $+5.302\text{e-05}$ \\
\hline
\end{tabular}
\caption*{Learned length 10 $h'$ and length 18 $h$ filters ($\lambda_3$ was increased to $8\text{e-4}$).}
\end{table}
\end{minipage}
\hfill
\begin{minipage}[c]{0.6\textwidth}
\includegraphics[width=1\textwidth]{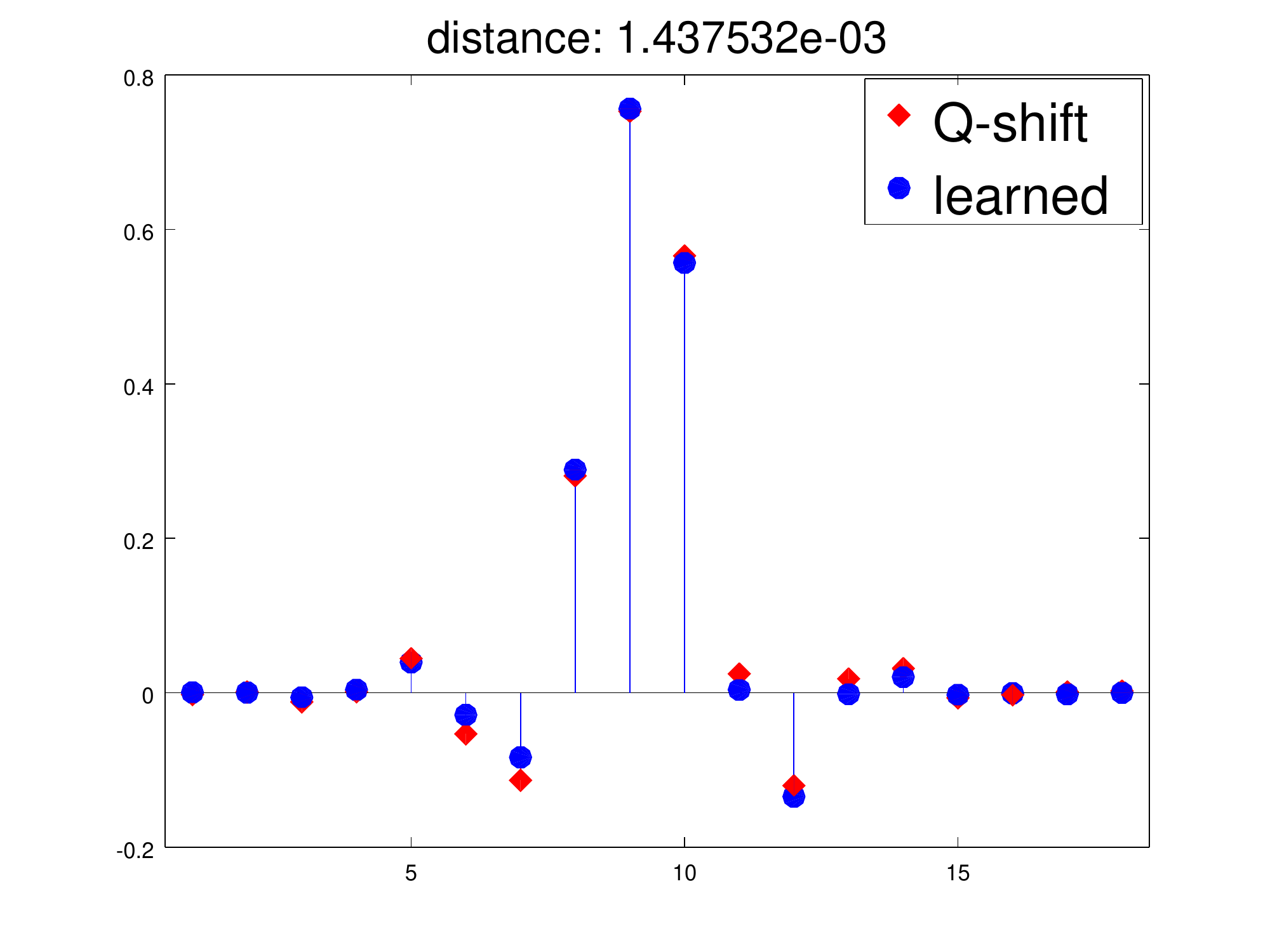}
\caption{Comparison of Kingsbury's Q-shift 18 tap filter with the learned length 18 $h$.}
\end{minipage}
\end{figure}

\section{Full Dual-tree Complex Wavelet Transform Network}
\begin{figure}[H]
\noindent\makebox[\textwidth]{%
\centering
\includegraphics[width=1.2\textwidth]{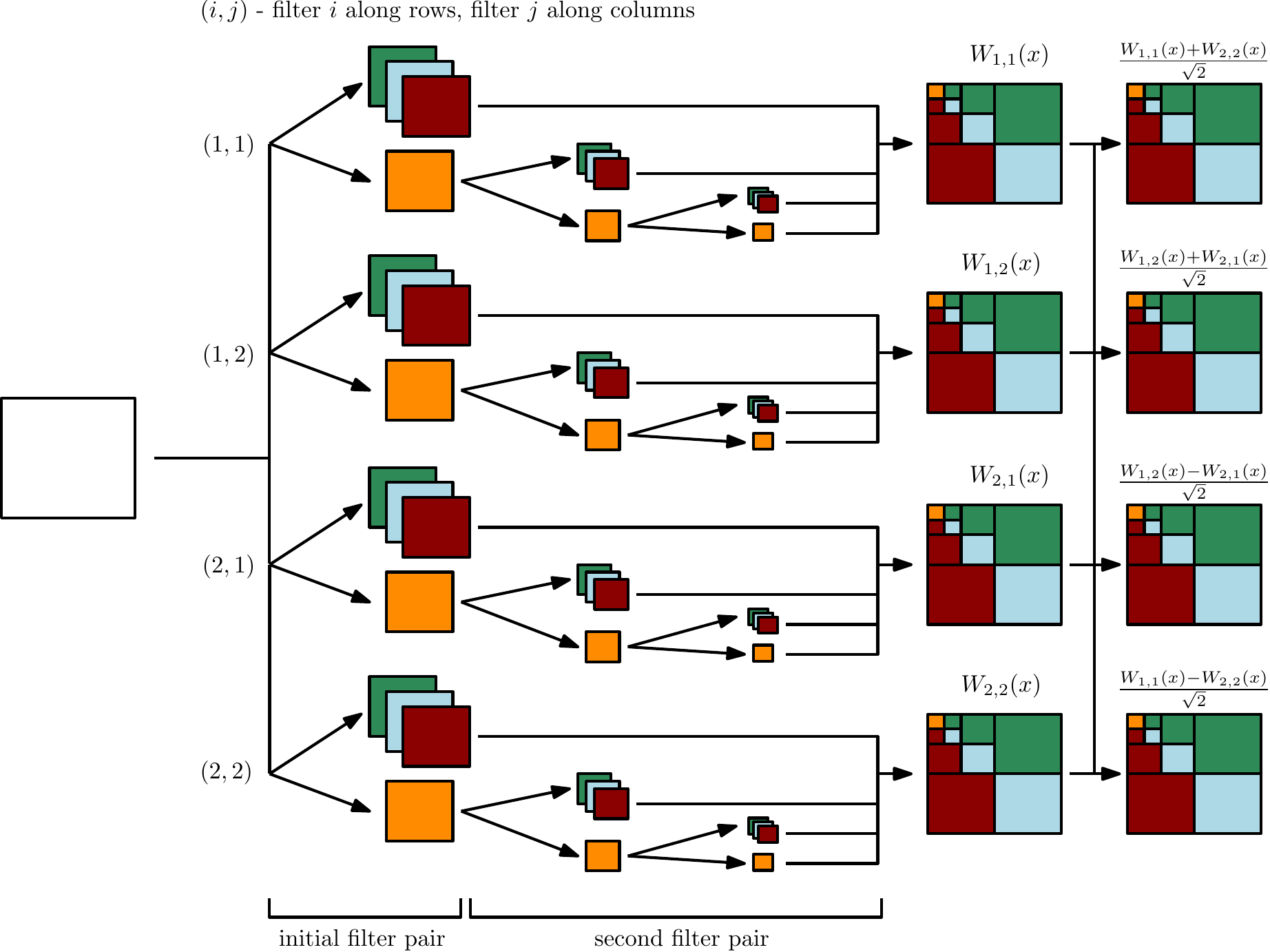}}
\caption{The full complex dual-tree wavelet transform network. Each $W_{i,j}$ is computed as in Equations \ref{eq:wsum1}--\ref{eq:wdiff2}.}
\label{fig:2dwnn-full}
\end{figure}

\end{appendices}


\begin{thebibliography}{10}

\bibitem{tensorflow-short}
Mart\'{\i}n Abadi et~al.
\newblock {TensorFlow}: Large-scale machine learning on heterogeneous systems,
  2015.
\newblock Software available from tensorflow.org.

\bibitem{bruna2011classification}
Joan Bruna and St{\'e}phane Mallat.
\newblock Classification with scattering operators.
\newblock In {\em Computer Vision and Pattern Recognition (CVPR), 2011 IEEE
  Conference on}, pages 1561--1566. IEEE, 2011.

\bibitem{bruna2013invariant}
Joan Bruna and St{\'e}phane Mallat.
\newblock Invariant scattering convolution networks.
\newblock {\em IEEE transactions on pattern analysis and machine intelligence},
  35(8):1872--1886, 2013.

\bibitem{christopoulos2000jpeg2000}
Charilaos Christopoulos, Athanassios Skodras, and Touradj Ebrahimi.
\newblock The jpeg2000 still image coding system: an overview.
\newblock {\em IEEE transactions on consumer electronics}, 46(4):1103--1127,
  2000.

\bibitem{haar1910theorie}
Alfred Haar.
\newblock Zur theorie der orthogonalen funktionensysteme.
\newblock {\em Mathematische Annalen}, 69(3):331--371, 1910.

\bibitem{hadji2017spatiotemporal}
Isma Hadji and Richard~P Wildes.
\newblock A spatiotemporal oriented energy network for dynamic texture
  recognition.
\newblock In {\em Proceedings of the IEEE Conference on Computer Vision and
  Pattern Recognition}, pages 3066--3074, 2017.

\bibitem{hinton2006reducing}
Geoffrey~E Hinton and Ruslan~R Salakhutdinov.
\newblock Reducing the dimensionality of data with neural networks.
\newblock {\em science}, 313(5786):504--507, 2006.

\bibitem{kingma2014adam}
Diederik Kingma and Jimmy Ba.
\newblock Adam: A method for stochastic optimization.
\newblock {\em arXiv preprint arXiv:1412.6980}, 2014.

\bibitem{kingsbury1998dual}
Nick Kingsbury.
\newblock The dual-tree complex wavelet transform: a new efficient tool for
  image restoration and enhancement.
\newblock In {\em Signal Processing Conference (EUSIPCO 1998), 9th European},
  pages 1--4. IEEE, 1998.

\bibitem{mallat}
St{\'e}phane Mallat.
\newblock {\em A Wavelet Tour of Signal Processing, Third Edition: The Sparse
  Way}.
\newblock Academic Press, 3rd edition, 2008.

\bibitem{mallat2012group}
St{\'e}phane Mallat.
\newblock Group invariant scattering.
\newblock {\em Communications on Pure and Applied Mathematics},
  65(10):1331--1398, 2012.

\bibitem{mallat2016understanding}
St{\'e}phane Mallat.
\newblock Understanding deep convolutional networks.
\newblock {\em Phil. Trans. R. Soc. A}, 374(2065):20150203, 2016.

\bibitem{mallat1989multiresolution}
Stephane~G Mallat.
\newblock Multiresolution approximations and wavelet orthonormal bases of
  ${L}^2({R})$.
\newblock {\em Transactions of the American mathematical society},
  315(1):69--87, 1989.

\bibitem{mallat1989theory}
Stephane~G Mallat.
\newblock A theory for multiresolution signal decomposition: the wavelet
  representation.
\newblock {\em IEEE transactions on pattern analysis and machine intelligence},
  11(7):674--693, 1989.

\bibitem{recoskie2018blearning}
Daniel Recoskie and Richard Mann.
\newblock Learning filters for the 2{D} wavelet transform.
\newblock In {\em Computer and Robot Vision (CRV), 2018 15th Conference on},
  pages 1--7. IEEE, 2018.

\bibitem{recoskie2018alearning}
Daniel Recoskie and Richard Mann.
\newblock Learning sparse wavelet representations.
\newblock {\em arXiv preprint arXiv:1802.02961}, 2018.

\bibitem{selesnick2005dual}
Ivan~W Selesnick, Richard~G Baraniuk, and Nick~C Kingsbury.
\newblock The dual-tree complex wavelet transform.
\newblock {\em IEEE signal processing magazine}, 22(6):123--151, 2005.

\bibitem{singh2017dual}
Amarjot Singh and Nick Kingsbury.
\newblock Dual-tree wavelet scattering network with parametric log
  transformation for object classification.
\newblock In {\em Acoustics, Speech and Signal Processing (ICASSP), 2017 IEEE
  International Conference on}, pages 2622--2626. IEEE, 2017.

\bibitem{singh2017efficient}
Amarjot Singh and Nick Kingsbury.
\newblock Efficient convolutional network learning using parametric log based
  dual-tree wavelet scatternet.
\newblock In {\em Proceedings of the IEEE Conference on Computer Vision and
  Pattern Recognition}, pages 1140--1147, 2017.

\end{thebibliography}
\end{document}